\documentclass[reprint,twocolumn,prx,aps,superscriptaddress,floatfix]{revtex4-2}
\usepackage{amsmath}
\usepackage{amssymb}
\usepackage{amsthm}
\usepackage[T1]{fontenc}
\usepackage[utf8]{inputenc}
\usepackage{amsfonts}
\usepackage{dsfont}
\usepackage{listings}
\usepackage{mathbbol}
\usepackage{physics}
\usepackage{enumerate}
\usepackage{latexsym}
\usepackage{psfrag}
\usepackage{bm}
\usepackage{graphicx}
\usepackage[caption=false]{subfig}
\usepackage{blkarray}
\usepackage{array}
\usepackage{color}
\usepackage[normalem]{ulem}
\usepackage{hyperref}
\usepackage{enumitem}
\usepackage{mathtools}
\hypersetup{
     colorlinks = true,
     linkcolor = magenta,
     citecolor = magenta
}

\def\beq{\begin{equation}}
\def\eeq{\end{equation}}
\def\bal{\begin{aligned}}
\def\eal{\end{aligned}}

\allowdisplaybreaks

\begin{document}

\begin{abstract}
Higher vortexability is often viewed as a route to topological flat bands with higher-Landau-level-like quantum geometry. Here we emphasize a complementary perspective: it provides a tunable multiband structure in which wave function geometry can be varied continuously while the band dispersion, degeneracy, and topology remain fixed. We perform systematic exact-diagonalization studies of many-body phases in fractionally filled higher vortexable moir\'e systems, retaining the full flat-band Hilbert space rather than projecting onto a single band. The multiband treatment reveals a cascade of Abelian and non-Abelian phases at zero magnetic field, including integer and fractional exciton insulators, Abelian fractional Chern insulators, Moore–Read and Read–Rezayi states. At fixed filling, different phases are connected through transitions or crossovers driven solely by changes in wave function geometry, highlighting quantum geometry itself as a direct tuning parameter between competing topological states. At fillings associated with Moore–Read and Read–Rezayi states, our calculations show that interband mixing shifts the optimal quantum geometry regime without suppressing non-Abelian topological order under screened Coulomb interaction. Our results establish higher vortexable moir\'e bands as a tunable platform for exploring geometry-driven multiband topological phases at zero magnetic field.
\end{abstract}
\title{Tunable Multiband Geometry and Fractional Phases in Higher Vortexable Systems}
\author{Xiaohan Wan}\thanks{These two authors contributed equally}
\affiliation{%
Department of Physics, University of Washington, Seattle, WA 98195, USA}
\author{Siddhartha Sarkar}\thanks{These two authors contributed equally}
\affiliation{Max Planck Institute for the Physics of Complex Systems, N\"othnitzer Stra\ss e 38, 01187 Dresden, Germany}
\author{Ting Cao}
\email{tingcao@uw.edu}
\affiliation{%
Department of Material Science and Engineering, University of Washington, Seattle, Washington 98195, USA}
\author{Mark Rudner}
\email{rudner@uw.edu}
\affiliation{%
Department of Physics, University of Washington, Seattle, WA 98195, USA}
\author{Di Xiao}
\email{dixiao@uw.edu}
\affiliation{%
Department of Physics, University of Washington, Seattle, WA 98195, USA}
\affiliation{%
Department of Material Science and Engineering, University of Washington, Seattle, Washington 98195, USA}
\author{Kai Sun}
\email{sunkai@umich.edu}
\affiliation{%
Department of Physics, University of Michigan, Ann Arbor, MI 48109, USA
}
\maketitle

\section{Introduction}
\label{sec:intro}
A central lesson from fractional quantum Hall and fractional quantum anomalous Hall physics~\cite{stormer1999fractional,laughlin1999nobel,tang2011high,sun2011nearly,neupert2011fractional,sheng2011fractional,regnault2011fractional,xiao2011interface,bernevig2012emergent,wu2012adiabatic,parameswaran2013fractional,roy2014band,wu2015fractional,repellin2020chern,ledwith2020fractional,simon2020contrasting,liu2021gate,mera2021engineering,li2021spontaneous,Devakul2021Magic,ledwith2021strong,wang2021exact,xie2021fractional,cai2023signatures,zeng2023thermodynamic,park2023observation,xu2023observation,ledwith2023vortexability,wu2024quantum,lu2024fractional,wang2024fractional,xie2025tunable} is that the fate of an interacting flat-band system is not determined by band dispersion alone. The topology and quantum geometry of the single particle wave functions are equally essential. When a lattice Chern band, or a moir\'e miniband, closely mimics the topology and geometry of the lowest Landau level (LLL), Coulomb interactions can stabilize fractional topological orders analogous to those realized in conventional fractional quantum Hall systems.

This analogy is powerful but far from automatic. Landau levels and lattice Chern bands have distinct microscopic origins, distinct symmetries, and distinct wave function structures: magnetic translations versus lattice translations, continuous rotations versus crystalline point-group symmetries, and Landau level wave functions versus Bloch wave functions. The fact that these systems can nevertheless support the same many-body topological orders points to a deeper organizing principle. Recent theoretical work has identified this principle in terms of quantum geometry, especially ideal quantum geometry and vortexability~\cite{parameswaran2013fractional,roy2014band,ledwith2020fractional,ledwith2021strong,wang2021exact,mera2021engineering,ledwith2022family,ledwith2023vortexability}.
In the ideal limit, the projected density operators of a topological flat band obey the same algebraic structure as those of the LLL, and the corresponding Bloch wave functions can be viewed as LLL wave functions dressed by a periodic modulation inherited from the lattice. This perspective has helped clarify why fractional states can emerge in moir\'e materials such as twisted transition-metal dichalcogenides,
whose flat-band wave functions may lie close to this ideal geometric limit~\cite{cai2023signatures,zeng2023thermodynamic,park2023observation,xu2023observation,wang2024fractional,li2021spontaneous,reddy2023fractional}.

A recent development extends this analogy beyond the LLL through the notion of higher vortexability~\cite{fujimoto2025higher,liu2025theory}. In such systems, multiple vortexable topological flat bands are brought into exact degeneracy and hybridized by a chiral-symmetry-preserving tunneling term. Remarkably, this hybridization can leave the bands exactly flat and degenerate even when the tunneling is strong. In the strong-hybridization regime, one of the degenerate bands can acquire quantum geometry closely resembling that of a higher Landau level, such as the $n=1$ Landau level. Higher vortexable moir\'e bands therefore offer a route to emulate higher Landau level physics in zero magnetic field and provide a new pathway toward non-Abelian fractional phases.

Most existing studies of higher vortexability have focused on precisely this single-band viewpoint: one projects onto the band whose geometry resembles a higher Landau level and studies the resulting interacting problem~\cite{liu2025non,liu2025parafermions}. This approach has yielded important insights, but it also discards a defining feature of the higher vortexable construction. Higher vortexability does not merely produce one flat band that imitates a higher Landau level. Instead, it naturally produces an exactly degenerate multiband manifold, only one member of which may resemble a higher Landau level in an appropriate limit. Ignoring the remaining degenerate flat bands therefore removes a central ingredient of the problem and may miss fractional phases with no direct analogue in a single Landau level.

In this work, we retain this full degenerate manifold and show that it gives rise to qualitatively new many-body physics beyond the single-band approximation. The higher vortexable setting provides two opportunities that are generally absent in ordinary Landau level systems. First, it realizes a strongly hybridized, exactly flat, exactly degenerate multiband structure, creating a platform for fractional phases beyond the conventional single Landau level paradigm. Second, the interlayer hybridization strength tunes the Bloch wave functions, and hence the quantum metric and related geometric data, while leaving the band dispersion, degeneracy, and topology unchanged. higher vortexable bands therefore provide a setting in which quantum geometry can be varied as an independent control parameter without altering single particle energetics, allowing one to isolate its role in stabilizing and selecting competing fractional phases. Our studies find a rich cascade of Abelian and non-Abelian fractional phases. We show that distinct topological states can be connected by phase transitions or crossovers, driven purely by changes in quantum geometry, with the single particle energetics held fixed. We further show that the multiband structure need not hinder stabilization of non-Abelian states under screened Coulomb interaction.

The remainder of this paper is organized as follows. Section~\ref{sec:summary} provides a brief review of the essential physics of higher vortexable bands and summarizes our main results. Section~\ref{sec:model} presents the model and exact-diagonalization setup. Section~\ref{sec:phases} maps out the many-body phases in the weak- and strong-tunneling limits at fillings $\nu=1$, $1/3$, $2/3$, $2/5$, and $8/5$, establishing higher vortexable moir\'e systems as a natural platform for Halperin-type states, including integer and fractional exciton insulators. Section~\ref{sec:transitions} studies how these phases are connected as the quantum geometry is tuned, revealing geometry-driven phase transitions and crossover at different fillings. Section~\ref{sec:nonabelian} revisits fillings associated with non-Abelian phases and shows that two-band exact diagonalization (ED) quantitatively shifts the optimal quantum geometry regime relative to the single-band approximation. Section~\ref{sec:discussion} discusses experimental implications.
\section{Overview of Higher Vortexability and Summary of Results}
\label{sec:summary}

In this section, we first provide a brief review of vortexable and higher vortexable bands, and then summarize the key results of our study.

\subsection{Vortexability}

A vortexable band is a lattice generalization of the LLL. Consider a Dirac particle in a magnetic field,
\begin{equation}
H_{\text{g}}=v_{\text{F}}\begin{pmatrix} 0 & \Pi \\ \overline{\Pi} & 0 \end{pmatrix},
\label{eq:diracb}
\end{equation}
where $\overline{\Pi}=\hbar(-2 i\overline{\partial}+\frac{e}{\hbar}\overline{A})$ is the momentum. Here, $\overline{\partial}=\tfrac{1}{2}(\partial_{x}+i\partial_{y})$, $\overline{A}=A_{x}+iA_{y}$, and $\mathbf{A}$ is the vector potential satisfying $\nabla \times \mathbf{A}=B \hat{z}$. The null space of $\overline{\Pi}$, defined by $\overline{\Pi}\psi(\mathbf{r})=0$, forms the LLL. Its wave functions have the analytic structure
\begin{equation}
\psi^{\mathrm{LLL}}(\mathbf{r})
=f(z)e^{-z\overline{z}/4\ell _B^{2}},
\end{equation}
where $f(z)$ is an arbitrary holomorphic function of the complex coordinate $z=x+iy$, and \(\ell_B = \sqrt{\hbar/eB}\) is the magnetic length.

The two factors in the LLL wave function play distinct roles. The holomorphic factor $f(z)$ is the central analytic ingredient in quantum Hall and fractional quantum Hall physics, as exemplified by the holomorphic polynomial structure of the Laughlin state. By contrast, the Gaussian envelope characterizes microscopic physics, such as the magnetic length. Thus, to generalize the essential Landau level structure to lattice systems, the key requirement is to preserve the holomorphic dependence rather than the precise Gaussian envelope. Vortexability provides a precise theoretical framework for implementing this idea in a generic setting.

To formulate this framework, we replace the Landau level operator $\overline{\Pi}$ by a generic operator $\mathcal D$ and consider the Hamiltonian
\begin{equation}
H=\begin{pmatrix} 0 & \mathcal{D}^{\dagger} \\ \mathcal{D} & 0 \end{pmatrix}.
\label{eq:chiralh}
\end{equation}
Here, we assume that $\mathcal{D}$ has a zero mode $g(z,\overline{z})$ satisfying
\begin{equation}
\mathcal{D}g(z,\overline{z})=0,
\end{equation}
and that $\mathcal{D}$ contains only anti-holomorphic derivatives $\overline{\partial}$ and no holomorphic derivatives $\partial$. Then, for any holomorphic function $f(z)$,
\begin{equation}
\mathcal{D}\left[f(z)g(z,\overline{z})\right]
=f(z)\mathcal{D}g(z,\overline{z})=0.
\end{equation}
This property—that any zero mode remains a zero mode after multiplication by an arbitrary holomorphic function—is known as \emph{vortexability}. As in the LLL, this holomorphic multiplication generates an extensive null space of $\mathcal{D}$, which forms a flat band, with wave functions factorizing into two parts,
\begin{equation}
\psi(\mathbf{r})=f(z)g(z,\overline{z}),
\end{equation}
where $f(z)$ carries the holomorphic Landau level structure and $g(z,\overline{z})$ is a fixed envelope function, replacing the Gaussian function of the LLL. Such a flat band is known as a vortexable band. In a periodic lattice or moir\'e system, the wave function of this flat band obeys Bloch's theorem and can therefore be written as
\begin{equation}\label{eq:idealbandwfgeneral}
\psi_{\mathbf{k}}(\mathbf{r})
= \psi_{\mathbf{k}}^{\mathrm{LLL}}(\mathbf{r}) h(\mathbf{r}),
\end{equation}
where $\mathbf{k}$ is the Bloch wavevector and
$\psi_{\mathbf{k}}^{\mathrm{LLL}}(\mathbf{r})$ is the LLL wave function on a torus with magnetic translation symmetry (see Appendix A for an explicit expression of $\psi_{\mathbf{k}}^{\mathrm{LLL}}(\mathbf{r})$). $h(\mathbf{r})$ is a $\mathbf{k}$-independent function. Thus, up to the fixed dressing factor $h(\mathbf r)$, a vortexable band has the same analytic wave function structure as the LLL. This provides a precise lattice generalization of LLL physics, where a vortexable band inherits key properties of the LLL, such as ideal quantum geometry and nontrivial topology. 

At the level of single particle wave functions, the distinction between the LLL and a vortexable band is entirely encoded in the dressing factor $h(\mathbf r)$. In the LLL, $h(\mathbf r)=1$ is spatially uniform and isotropic, reflecting the continuous spatial symmetries of LLs. In a moir\'e lattice, by contrast, $|h(\mathbf r)|$ is a periodic function over the moir\'e unit cell~\cite{ledwith2020fractional,wang2021exact,fujimoto2025higher,sarkar2025unconventional,zhang2025beyond}, imprinting periodic density modulations over the moir\'e lattice and reducing the continuous spatial symmetry to the discrete point- and space-group symmetries of the moir\'e lattice. 

We conclude this discussion by presenting, as an example, a simple and widely used class of vortexable flat-band models~\cite{sarkar2023symmetry}. Consider a chiral operator of the form
\begin{equation}
\mathcal{D}
=(-2i\overline{\partial})^w \mathds{1}
+D_U(\mathbf{r};\boldsymbol{\alpha}),
\end{equation}
where $(-2i\overline{\partial})^w$ is the kinetic operator of the parent system before the moir\'e structure is introduced, and the integer $w$ is the winding number of a symmetry-protected band crossing: $w=1$ corresponds to a Dirac crossing, while $w=2$ corresponds to a quadratic band crossing point. The periodic function $D_U(\mathbf{r};\boldsymbol{\alpha})$ represents the moir\'e potential or tunneling~\cite{tarnopolsky2019origin,li2022magic,becker2022fine,le2022double,eugenio2022twisted,becker2023degenerate,wan2023topological}, with $\boldsymbol{\alpha}$ denoting the corresponding tuning parameters (e.g., twisting angle). The crucial feature of this $\mathcal{D}$ is that it contains only anti-holomorphic derivatives $\overline{\partial}$ and no holomorphic derivatives $\partial$, allowing $\mathcal{D}$ to support vortexable flat bands. In fact, many commonly studied chiral moir\'e models, including chiral twisted bilayer graphene, fall into this class.  Note that since $\psi_\mathbf{k}^\text{LLL}$ must have one zero per (magnetic) unit cell, the wave function $\psi_\mathbf{k}$ in Eq.~\eqref{eq:idealbandwfgeneral} must also have one zero per (moir\'e) unit cell. This zero appears at some ``magic'' values of the tuning parameters $\boldsymbol{\alpha}$~\cite{tarnopolsky2019origin,wan2023topological} resulting in flat vortexable Chern bands with ideal quantum geometry, providing a natural setting for fractional topological states~\cite{roy2014band,ledwith2020fractional}.

\subsection{Higher Vortexability}
While vortexability enables generalized LLL wave functions, many fractional phases of interest, including non-Abelian states associated with higher Landau levels, require a framework for generalizing higher Landau level wave functions to lattice systems. Higher vortexability was introduced precisely for this purpose.

The central idea can be illustrated using Bernal bilayer graphene in a magnetic field. Its Hamiltonian has the same structure as Eq.~\eqref{eq:chiralh}, with
\begin{align}
\mathcal{D}_{\text{bg}}=\begin{pmatrix} v_F\bar{\Pi} & \gamma_{\text{bg}} \\ 0 & v_F\bar{\Pi} \end{pmatrix},
\label{eq:Bernalbg}
\end{align} 
where $\gamma_{\text{bg}}$ is the interlayer tunneling strength. The operator $\mathcal{D}_{\text{bg}}$ has two types of zero modes, giving rise to two degenerate flat bands~\cite{mccann2013electronic},
\begin{equation}
\Phi_1=\left(\psi_{\mathbf{k}}^{\text{LLL}},0\right),\qquad
\Phi_2=\left(\frac{\ell _B}{\sqrt{2}}\psi_{\mathbf{k}}^{\text{LL1}}, -\frac{\hbar v_{\text{F}}}{\gamma_{\text{bg}}}\psi_{\mathbf{k}}^{\text{LLL}}\right)
\end{equation}
where $\psi_{\mathbf{k}}^{\text{LL1}}$ is the wave function of the $n=1$ Landau level (see Appendix A for an explicit expression of $\psi_{\mathbf{k}}^{\mathrm{LL1}}(\mathbf{r})$).

This construction shows how higher Landau level structure can emerge from coupled LLL building blocks. A single layer governed by $\overline{\Pi}$ has only LLL zero modes. However, when two such layers are coupled through the chiral block $\mathcal{D}_{\text{bg}}$, the zero-mode manifold contains not only the LLL-like state $\Phi_1$, but also a second state $\Phi_2$ whose first component carries first Landau level character. In the strong-tunneling limit $\gamma_{\text{bg}}\to\infty$, the LLL component of $\Phi_2$ vanishes, and $\Phi_2$ reduces to a pure first Landau level wave function.

Remarkably, this mechanism does not rely on the literal Landau level operator $\overline{\Pi}$. It can be generalized to lattice or moir\'e systems by replacing each $\overline{\Pi}$ block with any vortexable operator discussed in the previous section. This provides a route to constructing exactly flat moir\'e bands whose wave functions and quantum geometry emulate higher Landau levels, thereby creating a natural platform for fractional phases beyond the LLL paradigm.

We now consider a moir\'e Hamiltonian with the same chiral structure as Eq.~\eqref{eq:chiralh}, but with the operator $\mathcal{D}$ chosen to have the block form
\begin{equation}
\mathcal{D}_\text{hv}=\begin{pmatrix} \mathcal{D}_\text{v} & \mathcal{D}_{\gamma} \\ 0 & \mathcal{D}_\text{v}\end{pmatrix}.
\end{equation}
This operator has the same block structure as the Bernal bilayer graphene operator in Eq.~\eqref{eq:Bernalbg}. It can be viewed as a coupled bilayer moir\'e system: each layer is governed by the vortexable chiral operator introduced above $\mathcal{D}_\text{v}=(-2i\overline{\partial})^w \mathds{1}
+D_U(\mathbf{r};\boldsymbol{\alpha})$, while $\mathcal{D}_{\gamma}(\mathbf{r}) =\gamma (2i\overline{\partial})^{w-1} \mathds{1}$ is the non-moir\'e-periodic interlayer tunneling with $\gamma$ its strength. Since $\mathcal{D}_\text{v}$ supports a vortexable flat band, its zero modes take the form of Eq.~\eqref{eq:idealbandwfgeneral}. In analogy with Bernal bilayer graphene, the block operator $\mathcal{D}_\text{hv}$ has two types of zero modes, giving rise to two exactly degenerate flat bands:
\begin{align}
\Psi_{\mathbf{k},1}(\mathbf{r})&=\{\psi^{\mathrm{LLL}}_{\mathbf{k}}(\mathbf{r}),0\}\,h(\mathbf{r}),\notag\\
\Psi_{\mathbf{k},2}(\mathbf{r})&=\{\sin\theta\,\psi^{\mathrm{LL1}}_{\mathbf{k}}(\mathbf{r}),\cos\theta\,\psi^{\mathrm{LLL}}_{\mathbf{k}}(\mathbf{r})\}\,h(\mathbf{r}).
\label{eq:hv_wavefunctions}
\end{align}
Recall that $\psi^{\mathrm{LLL}}_{\mathbf{k}}(\mathbf{r})$ and $\psi^{\mathrm{LL1}}_{\mathbf{k}}(\mathbf{r})$ denote the normalized LLL and first Landau level wave functions on a torus, respectively. We have parameterized the interlayer coupling by an angle $\theta$ through $\tan\theta\propto\gamma$, which maps $\gamma\in[0,\infty)$ to $\theta\in[0,\pi/2)$ (the exact definition of $\theta$ is discussed in Sec.~\ref{sec:model}). Because of the multiplicative factor $h(\mathbf{r})$, $\Psi_{\mathbf{k},1}$ and $\Psi_{\mathbf{k},2}$ are independent but not generally orthogonal to each other. We  denote the orthonormalized wave functions by $\tilde{\Psi}_{\mathbf{k},1}$ and $\tilde{\Psi}_{\mathbf{k},2}$ (see Sec.~\ref{sec:model} for details). As $\theta$ increases from $0$ to $\pi/2$, the wave function of $\tilde{\Psi}_{\mathbf{k},2}$ smoothly evolves from LLL-like to first-Landau-level-like, while $\tilde{\Psi}_{\mathbf{k},1}$ remains LLL-like throughout, as shown in Fig.~\ref{fig:phasediagram}(a). This continuous tunability of quantum geometry, with dispersion, degeneracy, and topology held fixed, is a distinctive feature of higher vortexable systems.

We conclude this section by presenting the formal definition of higher vortexability~\cite{fujimoto2025higher}: a Chern band is higher vortexable if two conditions hold: (1) there exists a vortexable partner band with which it forms a vortexable pair; (2) the combined two-band subspace cannot be decomposed into two bands that are each vortexable on their own. Flat higher vortexable moir\'e bands have since been found in many models~\cite{fujimoto2025higher,liu2025theory,wan2023nearly,zhang2025beyond}, all realized in Hamiltonians of the same chiral form as Eq.~\eqref{eq:chiralh}.

\subsection{Summary of results}
\begin{figure}[t]
\centering
\includegraphics[width=\columnwidth]{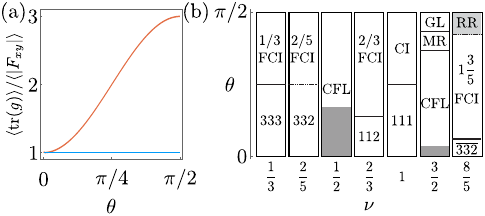}
\caption{Phase diagram of a moir\'e system with higher vortexable flat bands. (a)~Averaged ratio between the trace of the quantum metric and the Berry curvature as a function of $\theta$, which parameterizes the interlayer tunneling strength. The blue and red curves correspond respectively to the LLL-type vortexable band (band $1$, $\tilde{\Psi}_{\mathbf{k},1}$) and the first vortexable band (band $2$, $\tilde{\Psi}_{\mathbf{k},2}$). (b)~Phase diagram as a function of filling fraction $\nu$ and interlayer tunneling $\theta$. The filling fraction is defined such that $\nu=2$ corresponds to completely filling the two bands polarized on one sublattice. The ground states at different $\theta$ and $\nu$ are labeled as follows: \(p/q\)~FCI denotes a fractional Chern insulator corresponding to a lowest Landau level fractional quantum Hall state at filling \(p/q\); \(1 \frac{3}{5}\)~FCI denotes a completely filled band \(1\) together with a \(3/5\)-filled band \(2\); integer triplet $mmn$ denotes a Halperin-$mmn$ state; an overline denotes the corresponding state of holes relative to $\nu=2$; CFL denotes a composite Fermi liquid; CI denotes a Chern insulator; MR denotes the Moore--Read state; RR denotes the Read--Rezayi state; GL denotes a gapless state. The gray shading marks parameter ranges with strong system-size dependence; in particular, the stability of the RR phase under the screened Coulomb interaction remains uncertain in the thermodynamic limit.}
\label{fig:phasediagram}
\end{figure}

Although higher vortexable bands were originally motivated by the goal of realizing higher Landau level physics and the associated fractional phases in lattice systems, the construction reviewed above has a broader consequence. As shown in Eq.~\eqref{eq:hv_wavefunctions}, it produces two exactly degenerate flat bands and provides a knob, $\theta$, that continuously tunes the quantum geometry of one band from LLL-like to first-Landau-level-like. Crucially, this tuning leaves the other single particle ingredients unchanged: the bands remain exactly flat, exactly degenerate, and their topology remains fixed.

Previous studies of many-body ground states in higher vortexable moir\'e systems have largely employed single-band reductions. In one approach, the interaction Hamiltonian is projected into the first vortexable band, \(\Psi_2\), while the \(\Psi_1\) band is neglected~\cite{liu2025non,chen2025robust,zhang2025beyond}. In another, the two nearly degenerate bands are first split at the Hartree--Fock (HF) level, after which the interaction is projected onto a single HF band~\cite{fujimoto2025higher, ahn2024non}. However, because the HF splitting is itself generated by the Coulomb interaction, it is generically comparable to the residual interaction scale. Thus, there is no controlled separation of energy scales that would justify a single-band projection. Interband scattering terms remain an essential part of the problem and can qualitatively affect the many-body ground state when the two bands are nearly degenerate~\cite{rezayi2017landau}. Related work on multiband systems, such as bilayer graphene in a strong magnetic field, have primarily focused on the role of externally induced band splittings in the ground state at filling fraction $1/3$~\cite{le2023competing}.

Motivated by these considerations, we perform a systematic numerical study that retains the full two-band Hilbert space and elucidates the many-body phases of higher vortexable moir\'e systems over a range of filling fractions. Figure~\ref{fig:phasediagram}(b) summarizes the cascade of Abelian and non-Abelian fractional phases uncovered in this work at zero magnetic field. We briefly highlight two main results.
\begin{itemize}
\item {\bf A rich phase diagram with quantum-geometry-driven transitions.} 
For screened Coulomb interaction, the system exhibits sharply distinct behavior in the weak- and strong-tunneling regimes, which are separated by phase transitions or crossovers, and together give rise to a remarkably rich phase diagram. In the weak-tunneling regime, $\theta\approx0$, the system behaves similarly to a quantum Hall bilayer and stabilizes Halperin-like states. By contrast, in the strong-tunneling regime, $\theta\approx\pi/2$, the system effectively behaves as a single layer quantum Hall system and, depending on the filling, stabilizes integer-quantum-Hall-like, Laughlin-like, Jain-like, Moore--Read-like, and Read--Rezayi-like states. Since varying $\theta$ changes only the quantum geometry of the topological flat bands while leaving the single particle energetics fixed, these results demonstrate that quantum geometry alone can drive transitions or crossovers between distinct topological phases.

\item {\bf Multiband structure need not hinder non-Abelian states.} At fillings associated with Moore–Read and Read–Rezayi states, retaining the full two-band Hilbert space need not hinder non-Abelian topological order. Comparison with single-band exact diagonalization shows that interband scattering shifts the optimal quantum geometry regime for stabilizing non-Abelian phases instead of eliminating them. Within finite size resolution, higher vortexable moir\'e systems can host non-Abelian states with many-body gaps of the same order as those of the corresponding first Landau level states under screened Coulomb interaction.
\end{itemize}

More specifically, in Sec.~\ref{sec:phases}, we map out the many-body ground states in the weak-tunneling, $\theta\to0$, and strong-tunneling, $\theta\to\pi/2$, limits at fillings $\nu=1$, $1/3$, $2/3$, $2/5$, and $8/5$. Near $\theta=0$ and for screened Coulomb interaction, we identify the ground states as the Halperin-$111$, Halperin-$333$, Halperin-$112$, and Halperin-$332$ states, together with the Halperin-$332$ state of holes at $\nu=8/5$. The Halperin-$111$ and Halperin-$333$ states are integer and fractional exciton insulators, respectively, whose exciton condensation is supported by the eigenvalue spectrum of the exciton correlation matrix. By contrast, in the strong-tunneling limit $\theta=\pi/2$, the corresponding ground states are a Chern insulator, a $1/3$ Laughlin-like fractional Chern insulator, the particle-hole conjugate of a $1/3$ fractional Chern insulator (FCI), the Jain $2/5$ state, and the non-Abelian Read--Rezayi state at $\nu=8/5$. Since van der Waals materials typically have weak interlayer tunneling, corresponding to $\theta\sim0$ in our parameterization, our results suggest that higher vortexable moir\'e systems constructed from coupled vortexable layers are naturally positioned near the regime where Halperin-type fractional states are favored. Such systems, including double twisted bilayer graphene and strained bilayer graphene, may therefore provide promising zero-field platforms for realizing integer and fractional exciton insulators.

One of the central results of this work is presented in Sec.~\ref{sec:transitions}, where we investigate how the weak- and strong-tunneling limits are connected as $\theta$ is increased. At $\nu=2/3$, the Halperin-$112$ state at small $\theta$ and the $2/3$ FCI at large $\theta$ have the same intrinsic Abelian topological order but are separated by a sharp transition near $\theta\approx0.4$. We find that the transition is enforced by threefold rotation symmetry: the two ground state manifolds carry different $C_3$ quantum numbers and therefore cannot be continuously connected as long as the crystalline symmetry is preserved. By contrast, at $\nu=2/5$, the Halperin-$332$ state evolves smoothly into the Jain $2/5$ state without a gap closing, consistent with the fact that these two states share the same Chern--Simons description in the absence of rotational symmetry. At \(\nu=8/5\), the many-body gap closes near \(\theta\approx0.1\). On the larger-\(\theta\) side of this gap closing, an intermediate fivefold quasidegenerate ground state manifold appears and is labeled as a \(1\frac{3}{5}\) FCI. Near \(\theta=\pi/2\), five additional states descend in energy and together with the intermediate five states form the tenfold Read--Rezayi ground state manifold. At $\nu=1$ and $\nu=1/3$, with layer-resolved interactions, the integer and fractional exciton insulators evolve into the corresponding Chern-insulator and fractional-Chern-insulator states through a sequence of level crossings, each of which transfers one electron between the two bands. These results show that quantum geometry can tune not only between phases with distinct intrinsic topological order, but also between symmetry-distinguished realizations of the same intrinsic order.

In Sec.~\ref{sec:nonabelian}, we focus on fillings associated with non-Abelian phases in the presence of interband scattering. For both the Moore--Read state at $\nu=3/2$ and the Read--Rezayi state at $\nu=8/5$, 
we show that the full two-band simulation shifts the optimal quantum geometry regime relative to the single-band approximation under identical screened Coulomb interaction. Within finite size resolution, the many-body gaps of these multiband non-Abelian states are of the same order as the corresponding states in the first Landau level under the same screened Coulomb interaction. These results demonstrate that the multiband structure of higher vortexable systems need not hinder exotic non-Abelian phases. 
\section{Model}
\label{sec:model}
Here, we use the $w=2$ higher vortexable model as a concrete setting to demonstrate the underlying physics, while noting that the qualitative features discussed below can be generalized to broader classes of higher vortexable systems. The model describes a two-dimensional homobilayer system in which each layer hosts a quadratic band crossing point at a time-reversal-invariant momentum, subject to a chiral-symmetric moir\'e-periodic strain field~\cite{wan2023nearly,zhang2025beyond}. The single particle Hamiltonian, written in the basis
$(|1,A\rangle,|2,A\rangle,|1,B\rangle,|2,B\rangle)$, takes the form of Eq.~\eqref{eq:chiralh} with
\begin{equation}
\mathcal{D}=\begin{pmatrix} -4\bar{\partial}^{2}+\tilde{A}(\mathbf{r}) & 2i\gamma\bar{\partial} \\ 0 & -4\bar{\partial}^{2}+\tilde{A}(\mathbf{r}) \end{pmatrix}.
\label{eq:D}
\end{equation}
Here $1,2$ label the layer indices and $A,B$ denote the two sublattices. The diagonal terms of $\mathcal{D}$ and $\mathcal{D}^{\dagger}$ represent intralayer coupling, while the off-diagonal term encodes momentum-dependent tunneling between the two layers with tunneling amplitude $\gamma$. Within each layer, the system hosts a chiral-symmetric quadratic band touching described by $\partial^{2}$ and $\bar{\partial}^{2}$. Both layers are subject to the same moir\'e-periodic strain potential $\tilde{A}(\mathbf{r})=A_{x}(\mathbf{r})+iA_{y}(\mathbf{r})$. We consider a $C_{6v}$-symmetric strain up to third harmonic,
\begin{align}
\tilde{A}(\mathbf{r})=&-\alpha\sum_{n=1}^{3}\Big[ e^{i(1-n)\phi}\cos(\mathbf{G}_{n}\!\cdot\!\mathbf{r}) \nonumber
+\beta e^{i(2-n)\phi}\\
&\cos((\mathbf{G}_{n}-\mathbf{G}_{n+1})\!\cdot\!\mathbf{r}) \nonumber+\eta e^{i(1-n)\phi}\cos(2\mathbf{G}_{n}\!\cdot\!\mathbf{r})\Big],
\label{eq:strain}
\end{align}
where $\phi=2\pi/3$, $\mathbf{G}_{1}=G(0,1)$, and $\mathbf{G}_{2,3}=G(\mp\sqrt{3}/2,-1/2)$ are reciprocal lattice vectors with magnitude $G=4\pi/(\sqrt{3}a)$, and $a$ is the moir\'e lattice constant. $\mathbf{G}_4\equiv\mathbf{G}_1$. Here $\alpha$ characterizes the strength of the first harmonic component of the strain field, while $\beta$ and $\eta$ denote the relative strengths of the second and third harmonic components compared to the first. 

For nonzero $\gamma$, the single particle Hamiltonian with Eq.~\eqref{eq:strain} possesses chiral symmetry \(\mathcal{S}\) and the six-fold rotation \(C_{6z}\) about the $z$-axis normal to the 2D plane and composite magnetic \(M_z\mathcal{T}\) symmetries ($M_z: z\rightarrow-z$ is a mirror and $\mathcal{T}$ is time reversal) of the magnetic layer group \(p6/m'\), represented by
\begin{subequations}
\begin{align}
&D(C_{6z})=
\left(-\frac{1}{2}\sigma_0+\frac{\sqrt{3}}{2}i\sigma_z\right)\otimes\left(\frac{1-\omega^2}{2}\tau_0+\frac{1+\omega^2}{2}\tau_z\right), \\
&D(\mathcal{S})=\sigma_{z}\otimes\tau_0,\\
&D(M_{z}\mathcal{T})=\sigma_{x}\otimes(-i\tau_{y}),
\end{align}
\label{eq:symmetries1}
\end{subequations}
that satisfy 
\begin{subequations}
\begin{align}
&D(C_{6z})H(\mathbf{r})D(C_{6z})^{-1}=H(C_{6z}\mathbf{r}),\\
&D(\mathcal{S})H(\mathbf{r})D(\mathcal{S})^{-1}=-H(\mathbf{r}),\\
&D(M_{z}\mathcal{T})H^{*}(\mathbf{r})D(M_{z}\mathcal{T})^{-1}=H(\mathbf{r}),
\end{align}
\label{eq:symmetries2}
\end{subequations}
where $\omega=e^{i2\pi/3}$, $\sigma_0$ and $\tau_0$ denote $2\times2$ identity matrices, and $\sigma_{i}$ and $\tau_{i}$ ($i=x,y,z$) are Pauli matrices; $\sigma$ ($\tau$) matrices act on the sublattice (layer) degrees of freedom. When $\gamma=0$, the two layers decouple, and the bilayer Hamiltonian acquires the monolayer symmetry $p6mm$~\cite{wan2023topological} together with time reversal $\mathcal{T}$ and an $M_{z}$ symmetry that exchanges the two layers resulting in nonmagnetic $p6/mmm1'$ layer group symmetry. 

At critical values of the moir\'e potential parameters, for example $(\alpha,\beta,\eta)\approx(2.19\,G^{2},-0.5,-0.45)$, the system hosts four exactly flat bands: two polarized on the $A$ sublattice, and two polarized on the $B$ sublattice. Explicitly, the two sublattice-$A$-polarized flat band wave functions at $E=0$ can be written as:
\begin{align}
\Psi_{\mathbf{k},1}(\mathbf{r})&=\{\psi^{\mathrm{LLL}}_{\mathbf{k}}(\mathbf{r}),0, 0,0\}\,h(\mathbf{r}),\notag\\
\Psi_{\mathbf{k},2}(\mathbf{r})&=\{\ell _B\psi^{\mathrm{LL1}}_{\mathbf{k}}(\mathbf{r})/\sqrt{8},\gamma^{-1} \psi^{\mathrm{LLL}}_{\mathbf{k}}(\mathbf{r}),0,0\}\,h(\mathbf{r}).
\label{eq:wavefunctions}
\end{align}
Here, $\psi_\mathbf{k}^\text{LLL}(\mathbf{r})$ and $\psi_\mathbf{k}^\text{LL1}(\mathbf{r})$ are, respectively, the $n=0$ and $n=1$ Landau level wave functions on the torus in the Landau gauge, and $\ell _B=3^{1/4}a/(2\sqrt{\pi})$. For convenience, we define $\tan \theta \equiv  \ell_B\gamma/\sqrt{8}$ so that $\theta=0$ and $\theta=\pi/2$ correspond to $\gamma=0$ and $\gamma \to \infty$, respectively. Then we can write the second flat band wave function as $\Psi_{\mathbf{k},2}(\mathbf{r}) = \{\sin\theta \psi^\text{LL1}_\mathbf{k}(\mathbf{r}),\cos\theta\psi^\text{LLL}_\mathbf{k}(\mathbf{r}),0,0\}h(\mathbf{r})$. The other two sublattice-$B$-polarized wave functions can be obtained using $M_z \mathcal{T}$ symmetry: $\Psi_{\mathbf{k},3}(\mathbf{r}) =\{0,0,0, \overline{\psi_{-\mathbf{k}}^\text{LLL}(\mathbf{r})}\}\overline{h(\mathbf{r})}$ and $\Psi_{\mathbf{k},4}(\mathbf{r}) = \{0,0,-\cos\theta\overline{\psi_{-\mathbf{k}}^\text{LLL}(\mathbf{r})},\sin\theta\overline{\psi_{-\mathbf{k}}^\text{LL1}(\mathbf{r})}\}\overline{h(\mathbf{r})}$. We describe the origin of the flat band wave functions at $E=0$ in Appendix~\ref{app:origin}.
Taken together, the two bands polarized on the same sublattice ($\Psi_{\mathbf{k},1},\Psi_{\mathbf{k},2}$) have $|C|=2$ and satisfy the ideal non-Abelian quantum geometry condition $\mathrm{Tr}(g^{ij}_{\alpha\beta}(\mathbf{k}))=|\mathrm{Tr}(F^{ij}_{xy}(\mathbf{k}))|$ ($i,j\in \{1,2\}$ are band indices and $\alpha,\beta \in\{x,y\}$), exactly as the lowest two Landau levels taken together, where $g^{ij}_{\alpha\beta}(\mathbf{k})$ is the non-Abelian quantum metric and $F^{ij}_{xy}(\mathbf{k})$ is the non-Abelian Berry curvature. In moir\'e systems, $|h(\mathbf{r})|$ is moir\'e periodic and strongly inhomogeneous, which would typically cause large variation in Berry curvature and quantum metric across the Brillouin zone. However, we designed the strain field $\tilde{A}(\mathbf{r})$ such that the two flat bands on each sublattice have an almost constant quantum geometric tensor across the Brillouin zone~\cite{sarkar2026similar}.

To study the interacting phases, we first note that at filling fraction $\nu=2$, exchange interactions drive the electrons to occupy the two bands polarized on one sublattice, analogous to Hund's rule; the resulting ground state is a $|C|=2$ Chern insulator. For $\nu<2$, we anticipate that the same Hund's rule applies and project the interaction into the two corresponding flat bands:
\begin{equation}
H_{\mathrm{int}}=\frac{1}{2\mathcal{A}}\sum_{\mathbf{q}}V(\mathbf{q})\,\rho(\mathbf{q})\,\rho(-\mathbf{q}),
\label{eq:Hint}
\end{equation}
where $\mathcal{A}$ is the system area and $V(\mathbf{q})/\mathcal{A}=4\pi U\tanh(dq)/(N_{s}\sqrt{3}qa)$ is the symmetric dual-gate screened Coulomb interaction. Here $U$ is the bare Coulomb energy between two particles separated by one moir\'e lattice constant $a$, $d$ is the distance from the moir\'e material to each gate, and $N_{s}$ is the number of moir\'e unit cells. The projected density operator is $\rho(\mathbf{q})=\sum_{\mathbf{k}}\sum_{i,j=1}^{2}\lambda^{ij}_{\mathbf{q}}(\mathbf{k})\,c^{\dagger}_{\mathbf{k},i}c_{\mathbf{k}+\mathbf{q},j}$, where $\lambda^{ij}_{\mathbf{q}}(\mathbf{k})=\langle u_{\mathbf{k},i}(\mathbf{r})|u_{\mathbf{k}+\mathbf{q},j}(\mathbf{r})\rangle$ is the form factor, $c^{\dagger}_{\mathbf{k},i}$ is the creation operator for Bloch state $\tilde{\Psi}_{\mathbf{k},i}$, and $u_{\mathbf{k},j}$ is the periodic part of the Bloch wave function $\tilde{\Psi}_{\mathbf{k},j}$. The orthonormalized wave functions are $\tilde{\Psi}_{\mathbf{k},1} = \mathcal{N}_{\mathbf{k},1}\Psi_{\mathbf{k},1}$ and 
$\tilde{\Psi}_{\mathbf{k},2} = \mathcal{N}_{\mathbf{k},2}(\Psi_{\mathbf{k},2} - 
\langle \tilde{\Psi}_{\mathbf{k},1} | \Psi_{\mathbf{k},2} \rangle 
\tilde{\Psi}_{\mathbf{k},1})$ with $\mathcal{N}_{\mathbf{k},i}$ the normalization factors. At $\theta=0$, the projected interacting Hamiltonian has an exact $U(2)$ symmetry. Away from $\theta=0$, interlayer tunneling breaks this symmetry, leaving only the total $U(1)$ charge-conservation symmetry. All ED results reported below use the screened Coulomb interaction unless otherwise noted. Although the results are shown for specific system sizes, we have verified that the identified ground state phases persist for other system sizes unless otherwise noted.
 
\section{Many-Body Phases in the Weak- and Strong-Tunneling Limits}
\label{sec:phases}
In this section we map out the many-body ground states obtained from exact diagonalization in the two limits of weak and strong interlayer tunneling, $\theta\approx0$ and $\theta=\pi/2$, at fillings $\nu=1$, $1/3$, $2/5$, $2/3$, and $8/5$, with each filling treated in a separate subsection. In the weak-tunneling limit, the ground states at all five fillings are Halperin-type states, while in the strong-tunneling limit, the system effectively behaves like a single layer quantum Hall system. 
We defer the discussion of even-denominator fillings to Section~\ref{sec:nonabelian}. 

\begin{figure}[h]
\centering
\includegraphics[width=\columnwidth]{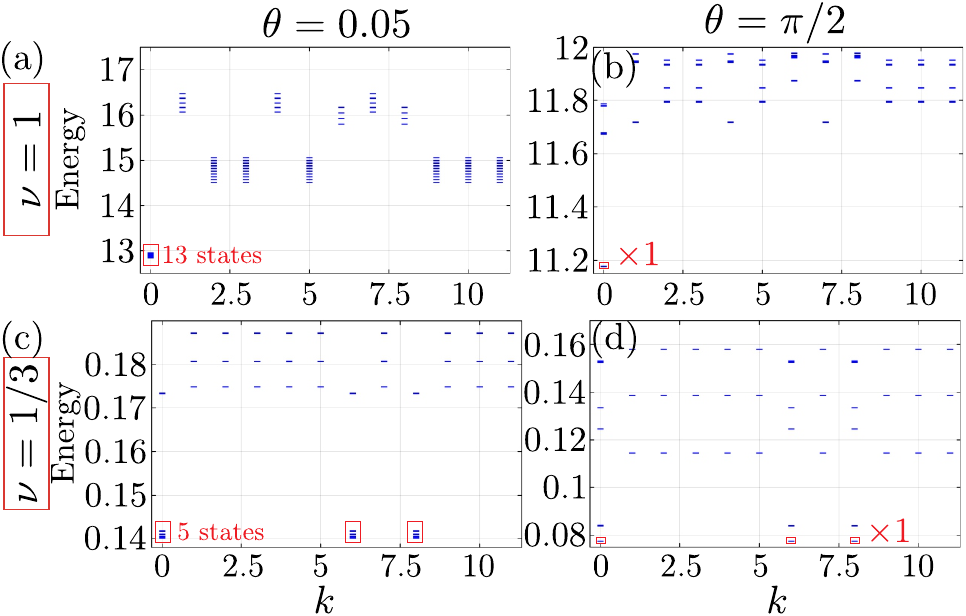}
\caption{Many-body spectra at fillings $\nu=1$ [(a),(b)] and $\nu=1/3$ [(c),(d)] for different values of the interlayer tunneling parameter $\theta$, with the ground states marked by red boxes. The average many-body Chern number of each ground state equals the filling fraction \(\nu\). (a)~At $\theta=0.05$, there are $13$ quasidegenerate ground states, consistent with a Halperin-$111$ state (exciton insulator). (b)~At $\theta=\pi/2$, there is a single ground state, adiabatically connected to a Chern insulator. (c)~At $\theta=0.05$, there are five quasidegenerate ground states in each of the three center of mass momentum sectors \(k=0,6,8\), giving a total of fifteen states, consistent with a Halperin-\(333\) state (fractional exciton insulator). (d)~At $\theta=\pi/2$, there are three quasidegenerate ground states, identified as a fractional Chern insulator with $C_{\mathrm{mb}}=1/3$ from the PES counting (Appendix~\ref{app:pes}).}
\label{fig:exciton}
\end{figure}

\subsection{\texorpdfstring{Exciton insulator and Chern insulator at $\nu=1$}{Exciton insulator and Chern insulator at nu=1}}
\label{sec:nu1}
\begin{figure}[t]
\centering
\includegraphics[width=\columnwidth]{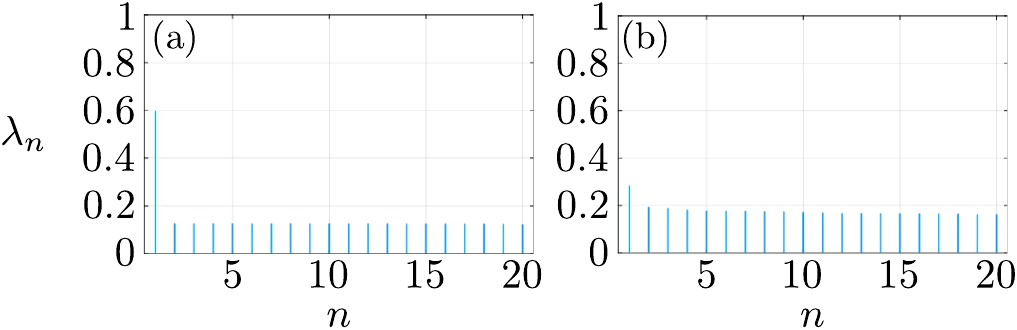}
\caption{Normalized eigenvalues of the exciton correlation matrix at $\nu=1,1/3$ for the $S_{z}=0$ ground state at $k=0$. (a)~$\nu=1$ and $\theta=0$, with eigenvalues normalized by $N_{e}/2=6$. (b)~$\nu=1/3$ and $\theta=0$, normalized by $N_{e}/2=2$.}
\label{fig:exciton_corr}
\end{figure}

The ED spectrum at $\nu=1$ for a system with $N_s = 12$ unit cells (see Appendix~\ref{app:clusters} for details of the system) and $\theta\approx 0$ is shown in Fig.~\ref{fig:exciton}(a). A manifold of $13$ ground states can be seen at center of mass (COM) momentum $k=0$ (see Appendix~\ref{app:clusters} for momentum ordering). The average many-body Chern number per ground state is $C_\text{mb}=1$. These properties are consistent with a Halperin-$111$ state, i.e., an exciton insulator in which the many-body wave function forms a ferromagnet in layer-pseudospin space. The degeneracy can be understood from the dimension of the \(SU(2)\) irreducible representation carried by the ferromagnetic ground state. At \(\theta=0\), the Hamiltonian is \(SU(2)\)-symmetric, and the fully polarized state is obtained from the completely symmetric combination of \(N_e=\nu N_s=12\) spin-\(1/2\) degrees of freedom. This symmetric product forms the total-spin \(S=N_e/2\) representation of \(SU(2)\), whose dimension is \(2S+1=N_e+1=13\). When \(\theta\) is increased slightly from zero, the weak breaking of \(SU(2)\) symmetry lifts the degeneracy among the \(13\) states, but they remain separated from higher energy states by the many-body gap, as shown
in Fig.~\ref{fig:exciton}(a). To characterize exciton condensation, we compute the normalized eigenvalues of the exciton two-body density matrix 
$\left\langle
c^{\dagger}_{\mathbf{k}_4,2}
c_{\mathbf{k}_3,1}
c^{\dagger}_{\mathbf{k}_2,1}
c_{\mathbf{k}_1,2}
\right\rangle$
for the \(S_z=0\) sector ground state at
\(\theta=0\)~\cite{sethi2023excitonic}. We find a single dominant normalized eigenvalue of approximately $0.6$, as shown in Fig.~\ref{fig:exciton_corr}(a), indicating excitonic superfluid order. 

At \(\theta=\pi/2\), we find a gapped singly degenerate ground state (as shown in
Fig.~\ref{fig:exciton}(b)) with many-body Chern number \(C_{\mathrm{mb}}=1\); hence the ground state is a correlated Chern insulator. Although in this state electrons are not fully polarized into a single band, we show in Appendix~\ref{app:adiabatic}
that this state can be adiabatically connected to a Chern insulator obtained
by completely filling band \(1\) without closing the many-body gap.

\subsection{\texorpdfstring{Fractional exciton insulator and $1/3$ FCI at $\nu=1/3$}{Fractional exciton insulator and 1/3 FCI at nu=1/3}}
\label{sec:nu13}
At \(\nu=1/3\) and small \(\theta\), ED for a system with \(N_s=12\) unit cells (Appendix~\ref{app:clusters}) reveals five nearly degenerate ground states in each of \(\Gamma\), \(K\), and \(K'\) momentum sectors ($k=0,6,8$) of the COM Brillouin zone, yielding a total ground state manifold of fifteen states [Fig.~\ref{fig:exciton}(c)]. The average many-body Chern number per ground state is $C_\text{mb}=1/3$. These properties are consistent with a Halperin-\(333\) state, i.e., a fractional exciton insulator. The fivefold counting within each momentum sector is analogous to the \(\nu=1\) case. For the \(N_s=12\) cluster, \(\nu=1/3\) corresponds to \(N_e=4\), giving total pseudospin \(S=N_e/2=2\) and multiplet dimension \(2S+1=5\). Together with the three momentum sectors, this gives the fifteen-state ground state manifold described above. Moreover, the spectrum of the exciton correlation matrix for the $S_{z}=0$ ground state at COM momentum $\Gamma$ at $\theta=0$ in Fig.~\ref{fig:exciton_corr}(b) shows a single dominant eigenvalue. This supports exciton condensation at this fractional filling. The largest eigenvalue at $\nu=1/3$ is less dominant than that at $\nu=1$. This is due to a much smaller number of excitons at $\nu=1/3$ for this system size. 

At \(\theta=\pi/2\), the many-body spectrum instead exhibits one ground state in each of the \(\Gamma\), \(K\), and \(K'\) momentum sectors of the COM Brillouin zone [Fig.~\ref{fig:exciton}(d)]. Each ground state carries many-body Chern number \(C_{\mathrm{mb}}=1/3\).
To establish that this ground state manifold realizes a Laughlin-type FCI, we compute its particle entanglement spectrum (PES)~\cite{li2008entanglement,regnault2011fractional}. We partition the \(N_e\) electrons into subsystems \(A\) and \(B\), containing \(N_A\) and \(N_B=N_e-N_A\) electrons, respectively, and trace over subsystem \(B\) to obtain
\(
\rho_A=\operatorname{Tr}_B \rho,
\qquad
\rho=
\frac{1}{n_{\mathrm{gs}}}
\sum_{i=1}^{n_{\mathrm{gs}}}
|\Psi^{(i)}\rangle\langle\Psi^{(i)}|.
\)
Here, \(|\Psi^{(i)}\rangle\) denotes the \(i\)-th many-body ground state, and \(n_{\mathrm{gs}}\) is the ground state degeneracy.
For a \(\nu=1/3\) Laughlin FCI, the generalized Pauli principle~\cite{haldane1991fractional, Haldane2006FermiLiquid,regnault2011fractional, bernevig2012emergent}---no more than one particle may occupy any three consecutive orbitals---predicts that the number of low-lying PES levels for a particle cut with \(N_A\) particles on a system of \(N_s\) unit cells is
\(
\mathcal{N}_{1/3}(N_s,N_A)
=
N_s
\frac{(N_s-2N_A-1)!}
{N_A!(N_s-3N_A)!}.
\)
For \(N_A=2\) and \(N_s=12\), we observe a PES gap above \(42\) levels, in agreement with \(\mathcal{N}_{1/3}(12,2)=42\); see Appendix~\ref{app:pes} for details. This counting confirms that the ground state manifold at \(\theta=\pi/2\) is a Laughlin-type \(\nu=1/3\) FCI.

\subsection{\texorpdfstring{Halperin-332 state and $2/5$ Jain state at $\nu=2/5$}{Halperin-332 state and 2/5 Jain state at nu=2/5}}
\label{sec:nu25}
At filling \(\nu=2/5\), exact diagonalization of a system with \(N_s=15\) unit cells reveals one ground state in each of the five COM momentum sectors \(k=0,1,2,3,4\), both near \(\theta=0\) and at \(\theta=\pi/2\). These states form a fivefold quasidegenerate ground state manifold with average many-body Chern number \(C_{\mathrm{mb}}=2/5\); see Figs.~\ref{fig:spectra_2325}(a) and (b). Details of the cluster geometry and momentum-sector convention are provided in Appendix~\ref{app:clusters}.

To identify the phase near \(\theta=0\), we compute the particle entanglement spectrum at \(\theta=0.05\). For an \(N_A=2\) particle partition of the \(N_s=15\) system, the PES exhibits gaps above \(315\) and \(405\) levels. These level countings agree with those of the Halperin-\(332\) state in a bilayer quantum Hall system at filling \(2/5\); see Fig.~\ref{fig:pes_nu25}. We therefore identify the ground state manifold at \(\theta=0.05\) as the Halperin-\(332\) state. Since the Halperin-\(332\) state is a pseudospin-\(SU(2)\) singlet, it contributes a single state to each COM momentum sector.

At \(\theta=\pi/2\), we identify the ground state manifold as a \(\nu=2/5\) FCI by demonstrating that it is adiabatically connected to a fully band-\(1\)-polarized \(2/5\) Jain state; see Appendix~\ref{app:adiabatic} for details.

\subsection{\texorpdfstring{Halperin-112 state and $2/3$ FCI at $\nu=2/3$}{Halperin-112 state and 2/3 FCI at nu=2/3}}
\label{sec:nu23}
\begin{figure}[t]
\centering
\includegraphics[width=\columnwidth]{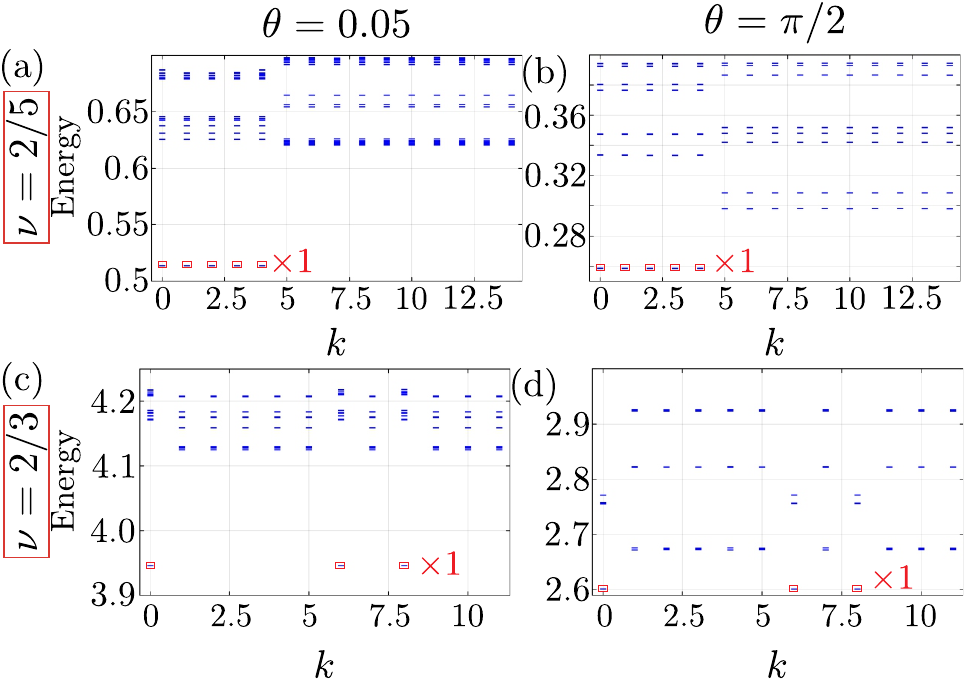}
\caption{Many-body spectra at $\nu=2/5$ [(a),(b)] and $\nu=2/3$ [(c),(d)] for $\theta=0.05$ and $\theta=\pi/2$.}
\label{fig:spectra_2325}
\end{figure}

At filling \(\nu=2/3\), ED of a system with \(N_s=12\) unit cells reveals one ground state in each of \(\Gamma\), \(K\), and \(K'\) momentum sectors ($k=0,6,8$) of the COM Brillouin zone. These states form a threefold ground state manifold with average many-body Chern number \(C_{\mathrm{mb}}=2/3\); see Fig.~\ref{fig:spectra_2325}.

To identify the phase near \(\theta=0\), we compute the particle entanglement spectrum at \(\theta=0.05\). For an \(N_A=3\) particle partition, the PES exhibits gaps above \(1520\) and \(1952\) levels [Fig.~\ref{fig:pes_nu23}(a)]. These level countings agree with those of the Halperin-\(112\) state in a bilayer quantum Hall system at filling \(2/3\), shown in Fig.~\ref{fig:pes_nu23}(b). We therefore identify the ground state manifold at \(\theta=0.05\) as the Halperin-\(112\) state.

At \(\theta=\pi/2\), we identify the ground state manifold as a \(\nu=2/3\) FCI using its hole particle entanglement spectrum. In the hole representation, this state may be viewed as having band \(2\) completely filled with holes and band \(1\) at hole filling \(1/3\). For a cut containing \(N_A=2\) holes on the \(N_s=12\) cluster, there are \(N_A+1=3\) possible distributions of holes between the two bands. When both holes occupy band \(2\), the ordinary Pauli principle gives
\(
\binom{12}{2}=66
\)
states. When one hole occupies each band, the additional contribution is
\(
\binom{12}{1}\mathcal{N}_{1/3}(12,1)
=
12\times 12
=
144,
\)
yielding a cumulative counting of
\(
66+144=210.
\)
Finally, when both holes occupy band \(1\), the generalized Pauli principle of the \(\nu=1/3\) Laughlin state gives
\(
\mathcal{N}_{1/3}(12,2)=42.
\)
The total counting is therefore
\(
66+144+42=252.
\)
The hole PES exhibits three gaps above \(66\), \(210\), and \(252\) levels, in agreement with this counting; see Appendix~\ref{app:pes}. These results confirm that the ground state manifold at \(\theta=\pi/2\) is a \(\nu=2/3\) FCI.

\subsection{\texorpdfstring{Halperin-332 state of holes and Read--Rezayi state at $\nu=8/5$}{Halperin-332 state of holes and Read--Rezayi state at nu=8/5}}
\label{sec:nu85_phases}
At \(\nu=8/5\) for a system with $N_s=15$ unit cells, we find a fivefold quasidegenerate ground state manifold at
small \(\theta\). The hole-PES gaps of these five states agree with the PES gaps of a Halperin-\(332\) state in a quantum Hall bilayer at \(2/5\)
filling; see Fig.~\ref{fig:pes_nu85_small} in Appendix~\ref{app:pes}. This
identifies the small-\(\theta\) ground state as a Halperin-\(332\) state of
holes.

Near \(\theta=\pi/2\), the many-body spectrum of an \(N_s=15\) cluster
contains two ground states in each of the five momentum sectors
\(k=0,1,2,3,4\), giving a total of ten states; see
Fig.~\ref{fig:nonabelian}(d). This tenfold degeneracy is consistent with the
expected ground state degeneracy of the Read--Rezayi
state. To confirm that the ground state at
\(\theta=\pi/2\) is a non-Abelian Read--Rezayi state, we compute its PES. Note that the ground state near \(\theta=\pi/2\) can be adiabatically connected to a state obtained by completely filling band \(1\) and partially filling band \(2\). If it is a Read--Rezayi (RR) state,
its PES counting can be obtained using the ordinary Pauli principle for band
\(1\) and the generalized Pauli principle of RR state in band \(2\), namely, no more
than three electrons may occupy any five consecutive orbitals. For a particle cut with \(N_A=4\), the expected total counting is obtained
by summing over $N_A+1=5$ different ways of distributing the electrons between
the two bands: (1) placing four electrons in band \(1\), which gives \(1365\)
states; (2) placing three electrons in band \(1\) and one electron in band
\(2\), which gives \(455\times15=6825\) states; (3) placing two electrons in
band \(1\) and two electrons in band \(2\), which gives
\(105\times105=11025\) states; (4) placing one electron in band \(1\) and
three electrons in band \(2\), which gives \(15\times455=6825\) states; and
(5) placing four electrons in band \(2\) subject to the generalized Pauli
principle---no more than three electrons within any five consecutive
orbitals---which gives \(1305\) states. Together, these contributions sum to
\(27345\) states.
In principle, the five ways of distributing the electrons between the two
bands could produce distinct groups of entanglement levels. If these groups
appear in the order listed above, for example, there could be gaps above
\(1365+6825\) states and above \(1365+6825+11025\) states. However, their
relative ordering in the PES and whether they are separated by visible gaps
depend on microscopic details of the form factors and interaction.
We observe only two gaps in the \(N_A=4\) PES of the ground state manifold
near \(\theta=\pi/2\), above \(1365\) and \(27345\) states, as shown in
Fig.~\ref{fig:pes_rr} of Appendix~\ref{app:pes}. The lowest \(1365\) levels
correspond to the possible configurations obtained by placing all four
electrons in band \(1\). The PES gap above \(27345\) states supports the
conclusion that the ten ground states near \(\theta=\pi/2\) form a
Read--Rezayi ground state manifold. We note that, for two-body interactions, finite size numerical studies of the \(Z_3\) Read--Rezayi phase often exhibit appreciable splitting of the topological ground state manifold and pronounced size dependence, while the stability of the phase can be sensitive to interaction parameters and competition with nearby charge-ordered states~\cite{zhu2015fractional,mong2017fibonacci,herviou2024numerical,liu2025parafermions}. Therefore, whether the RR state in the large-\(\theta\) limit survives under the screened Coulomb interaction in the thermodynamic limit remains unclear. Accordingly, we shade the region corresponding to the RR state in Fig.~\ref{fig:phasediagram}.

Since in van der Waals materials the interlayer tunneling term is typically small~\cite{fujimoto2025higher}, our results suggest higher vortexable moir\'e systems such as strained bilayer graphene and double twisted bilayer graphene as promising platforms for realizing Halperin-type states, including exciton insulators at zero magnetic field.
\section{Quantum-Geometry-Driven Phase Transitions and Crossovers}
\label{sec:transitions}
Having identified the weak- and strong-tunneling ground states, we now study how these phases are connected as $\theta$ is varied.

\subsection{Phase transition at $\nu=2/3$}
\label{sec:transition_nu23}
\begin{figure}[h]
\centering
\includegraphics[width=\columnwidth]{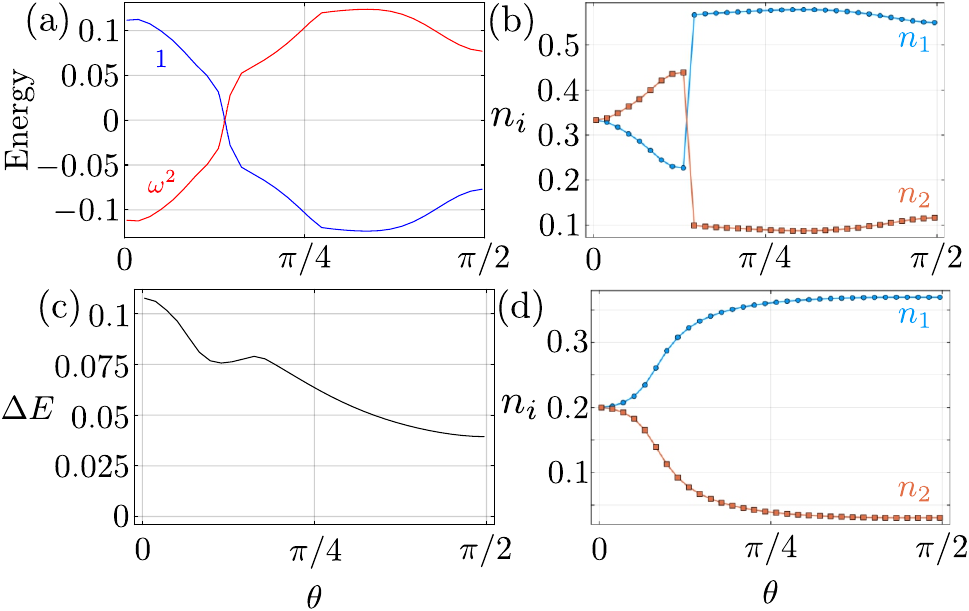}
\caption{Quantum-geometry-driven phase transition at $\nu=2/3$ [(a),(b)] and crossover at $\nu=2/5$ [(c),(d)]. (a)~Calibrated energies of the ground state and first excited state at $\mathbf{k}=0$ as functions of $\theta$; the average of the two energies is subtracted. A many-body level crossing occurs near $\theta=0.4$, where the $C_{3}$ eigenvalue of the ground state at $\mathbf{k}=0$ changes from $\omega^{2}$ to $1$. (b)  Ground state band occupation numbers at $\nu=2/3$ in band $\tilde{\Psi}_{\mathbf{k},1}$ ($n_{1}$, blue) and the first vortexable band $\tilde{\Psi}_{\mathbf{k},2}$ ($n_{2}$, red dotted), exhibiting a discontinuous jump at the band inversion. $n_i=\tfrac{1}{N_s}\sum_{\mathbf{k}}\langle c^\dagger_{\mathbf{k},i}c_{\mathbf{k},i}\rangle$. (c)~Many-body gap $\Delta E=E_{6}-E_{5}$ as a function of $\theta$ at $\nu=2/5$; the gap does not close over the full range. (d) Ground state occupation numbers at $\nu=2/5$.}
\label{fig:transition}
\end{figure}

Since the Halperin-$112$ state and the $2/3$ FCI share the same Abelian topological order~\cite{wen2004quantum,wen1992classification}, a smooth crossover is allowed from the perspective of topological field theory. However, our simulations indicate that they are separated by a phase transition. As $\theta$ is increased, the ground state undergoes an exact level crossing near $\theta\approx0.4$, as shown in Fig.~\ref{fig:transition}(a). This crossing is protected by rotational symmetry: the Halperin-$112$ state has $C_{3}$ eigenvalues $(\omega^{2},\omega,\omega)$ at $k=(0,6,8)$, whereas the $2/3$ FCI has corresponding $C_{3}$ eigenvalues $(1,\omega^{2},\omega^{2})$. Thus, although the two states are not distinguished by their Abelian topological order, they transform differently under $C_{3}$ rotation. Different $C_3$ eigenvalues of the two topological states result in an exact level crossing. At the crossing, the band occupation numbers exhibit a discontinuous jump as shown in Fig.~\ref{fig:transition}(b).

\subsection{\texorpdfstring{Crossover at $\nu=2/5$}
{Crossover at nu=2/5}}
\label{sec:transition_nu25}

At \(\nu=2/5\), the Halperin-\(332\) state at small \(\theta\) and the
\(2/5\) FCI at \(\theta=\pi/2\), identified in Sec.~\ref{sec:nu25}, are
allowed to be adiabatically connected because they share the same Abelian
topological order. On the finite size cluster we studied, the many-body
gap remains open as \(\theta\) is varied, as shown in
Fig.~\ref{fig:transition}(c), while the band occupation numbers evolve
smoothly, as shown in Fig.~\ref{fig:transition}(d). Since this cluster does
not possess rotational symmetry, there is no rotational-symmetry
quantum number that enforces an exact level crossing analogous to that at
\(\nu=2/3\). These finite size results suggest that Halperin-\(332\) state and \(2/5\) FCI are adiabatically connected, consistent with expectations from the corresponding $K$-matrix topological field theories.
\subsection{\texorpdfstring{From the Halperin-\(332\) state of holes to the
Read--Rezayi state at \(\nu=8/5\)}
{From the Halperin-332 state of holes to the Read--Rezayi state at nu=8/5}}
\label{sec:transition_nu85}
At \(\nu=8/5\), the many-body gap closes near
\(\theta\approx0.1\), as shown in Fig.~\ref{fig:gapclose_nu85}(a) in the Appendix. On the larger-\(\theta\) side of this gap closing, the many-body spectrum contains a fivefold
quasidegenerate ground state manifold, with one state in each of the momentum sectors \(k=0,1,2,3,4\). To determine the nature of the intermediate-\(\theta\) phase, we compute the hole PES of the ground state manifold at $\theta=0.25$, using a hole cut with $N_A=3$ and $N_s=15$. The hole PES exhibits a gap above \(350\) states, which is also present in the PES of a \(2/5\) Jain state in the LLL; see Fig.~\ref{fig:pes_nu85_large}. In addition, the
band occupation numbers show that the holes are predominantly polarized into band \(2\), as shown in Fig.~\ref{fig:gapclose_nu85}(b). These results suggest that the intermediate phase may be a band-$2$-polarized \(2/5\) Jain FCI of holes. However, the many-body gap closing near \(\theta\approx0.1\) shows that the intermediate
state is not smoothly connected, along the path obtained by varying \(\theta\), to the small-\(\theta\) \(\overline{332}\) state. This differs from the behavior at \(\nu=2/5\), where the Halperin-\(332\) state and the band-\(1\)-polarized \(2/5\) Jain FCI are adiabatically connected as \(\theta\) is varied. Since this phase corresponds, in terms of electrons, to a completely filled band \(1\) and a \(3/5\)-filled band \(2\), we label it as a \(1\frac{3}{5}\) FCI in the phase diagram. 

In the finite size spectrum, as \(\theta\) increases from \(0.25\) toward \(\pi/2\), the five intermediate-\(\theta\) ground states remain among the lowest-energy states. However, interestingly, close to \(\theta=\pi/2\), an additional set of five states descends in energy, and together the two quintets form the tenfold Read--Rezayi ground state manifold identified in Sec.~\ref{sec:nu85_phases}. This evolution is consistent with the Tao-Thouless thin-torus~\cite{tao1983fractional} structure of the two phases. The \(\nu=3/5\) Jain state has the root pattern~\cite{bergholtz2008quantum}
\(
11010\,11010\cdots
\) and its translations, while the fermionic \(\mathbb{Z}_3\) Read--Rezayi state has two sets of thin-torus ground state patterns~\cite{read1999beyond},
\(
11100\,11100\cdots
\qquad\text{and}\qquad
11010\,11010\cdots ,
\)
together with their translations. Thus, the Read--Rezayi manifold contains a quintet with Jain-like thin-torus structure and an additional quintet associated with the non-Abelian phase.

\subsection{Phase transitions at $\nu=1$ and $\nu=1/3$}
\label{sec:transition_nu1}
\begin{figure}[b]
\centering
\includegraphics[width=\columnwidth]{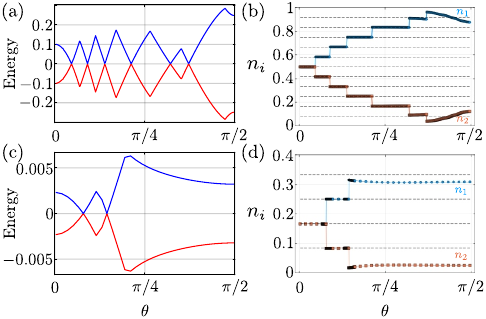}
\caption{Phase transitions at $\nu=1$ and $\nu=1/3$ for unequal intralayer and interlayer interactions. (a)~Energies of the two lowest states at $k=0$ versus $\theta$ at $\nu=1$, after subtracting their mean; both energies vanish at a level crossing. (b) Band-$i$ occupation number of the ground state versus $\theta$ at $\nu=1$. (c),(d)~Same quantities at $\nu=1/3$. In (d), the occupation numbers at each $\theta$ are averaged over the three ground states. The $N_s=12$ cluster is used, with $r=0$ for $\nu=1$ and $r=0.8$ for $\nu=1/3$.}
\label{fig:bandinversion}
\end{figure}
To study the transitions between a (fractional) exciton insulator and a (fractional) Chern insulator within finite size resolution, we consider a layer-resolved interaction Hamiltonian with different intralayer and interlayer interactions:
\begin{align}
\tilde{H}_{\mathrm{int}}=&\frac{1}{4A}\sum_{\mathbf{q}}\Big[\big(V_{\mathrm{intra}}(\mathbf{q})+V_{\mathrm{inter}}(\mathbf{q})\big)\big(\rho_{1}(\mathbf{q})+\rho_{2}(\mathbf{q})\big)\nonumber\\
&\big(\rho_{1}(-\mathbf{q})+\rho_{2}(-\mathbf{q})\big)+\big(V_{\mathrm{intra}}(\mathbf{q})-V_{\mathrm{inter}}(\mathbf{q})\big)\nonumber\\
&\big(\rho_{1}(\mathbf{q})-\rho_{2}(\mathbf{q})\big)\big(\rho_{1}(-\mathbf{q})-\rho_{2}(-\mathbf{q})\big)\Big],
\label{eq:Hint_modified}
\end{align}
where the layer-resolved projected density operator is $\rho_{l}(\mathbf{q})=\sum_{\mathbf{k}}\sum_{i,j=1}^{2}\langle u_{\mathbf{k},i}(\mathbf{r})|\hat{P}_{l}|u_{\mathbf{k}+\mathbf{q},j}(\mathbf{r})\rangle c^{\dagger}_{\mathbf{k},i}c_{\mathbf{k}+\mathbf{q},j}$. The operator $\hat{P}_{l}$ projects onto layer $l$. The first term describes a total density interaction, whereas the second term is analogous to the capacitance energy of a quantum Hall bilayer~\cite{ezawa2013quantum}. In the limit $V_\text{inter}=V_\text{intra}$, $\tilde{H}_{\mathrm{int}}$ reduces to $H_{\mathrm{int}}$ of Eq.~\eqref{eq:Hint}. 
At $\theta=0$, the first term is  $SU(2)$ invariant, while the second term is only layer-$U(1)\times U(1)$ invariant. For $\theta\neq0$, interlayer tunneling breaks the separate charge conservation in each layer, so the $U(1)\times U(1)$ symmetry is absent. We used $V_{\mathrm{intra}}(\mathbf{q})=V(\mathbf{q})$ and $V_{\mathrm{inter}}(\mathbf{q})=r\,V(\mathbf{q})+(1-r)\,4\pi U d/(N_{s}\sqrt{3}a)$ in numerical calculations. At $r=0$, $V_{\mathrm{inter}}$ is a contact interaction; at $r=1$, $V_{\mathrm{inter}}$ is the same as the intralayer screened Coulomb interaction. At $\theta=0$ and $V_{\text{inter}}\neq V_{\text{intra}}$, the degeneracy between the different $S_z$ sectors of the pseudospin ferromagnet (see Secs.~\ref{sec:nu1} and~\ref{sec:nu13}) is lifted due to broken $SU(2)$; in the thermodynamic limit,  states in different $S_z$ sectors form a continuous spectrum with $E\propto S_z^2$, whereas in finite size ED, the capacitance term separates the $S_z=0$ sector from states with larger $|S_z|$. At $\theta=\pi/2$, the ground state of $\tilde{H}_{\text{int}}$ remains the same as that of $H_{\text{int}}$ because $\langle\rho_2(\mathbf{q})\rangle\approx0$ (the electrons are polarized to band $1$). For $\tilde{H}_{\mathrm{int}}$, the two limits have the same number of ground states in a finite system, allowing us to track the transitions between the (fractional) exciton insulators and (fractional) Chern insulators. For the $C_6$-symmetric $N_s=12$ cluster, we find six level crossings at $\nu=1$ as $\theta$ is varied from $0$ to $\pi/2$, as shown in Fig.~\ref{fig:bandinversion}(a). The number of level crossings is dictated by the maximum number of excitons, $N_e/2$, which equals $6$ at $\nu=1$. Each level crossing corresponds to the transfer of one electron from band $2$ to band $1$, as shown in Fig.~\ref{fig:bandinversion}(b). Accordingly, the $C_6$ eigenvalue of the ground state at $k=0$ changes at each level crossing. Similarly, at $\nu=1/3$, we find $2$ level crossings over the full range $0\leq\theta<\pi/2$, as shown in Fig.~\ref{fig:bandinversion}(c). This number again equals the maximum number of excitons, $N_e/2=2$, at $\nu=1/3$. After the last level crossing at each filling, the band occupation numbers begin to vary continuously with $\theta$. The ground states at $\nu=1$ and $1/3$ evolve into an unpolarized Chern insulator and an unpolarized fractional Chern insulator, respectively.

Since varying $\theta$ changes only the Bloch wave functions while leaving the band dispersion, degeneracy, and topology fixed, the phase transitions and crossover studied in this section are driven purely by the change in quantum geometry of the first vortexable band (band $2$). This mechanism is distinct from that considered in previous studies of transitions between Halperin states and single layer fractional states, where an energy imbalance between the two layers is used as the tuning parameter~\cite{milovanovic2010transition}. Our results show that quantum geometry alone can induce a phase transition even when the single particle band dispersion, degeneracy, and topology remain fixed.
\section{Non-Abelian States: Two-Band versus Single-Band Exact Diagonalization}
\label{sec:nonabelian}
We now revisit the filling fractions associated with non-Abelian states, retaining the full two-band structure in exact diagonalization. A central message of this section is that multiband structure need not hinder the stabilization of non-Abelian states under screened Coulomb interaction. Compared with single-band exact diagonalization, two-band calculations quantitatively shift the optimal quantum geometry regime in which these states are stabilized.

\subsection{\texorpdfstring{Moore--Read state at $\nu=3/2$}{Moore--Read state at nu=3/2}}
\label{sec:nu32}
\begin{figure}[t]
\centering
\includegraphics[width=\columnwidth]{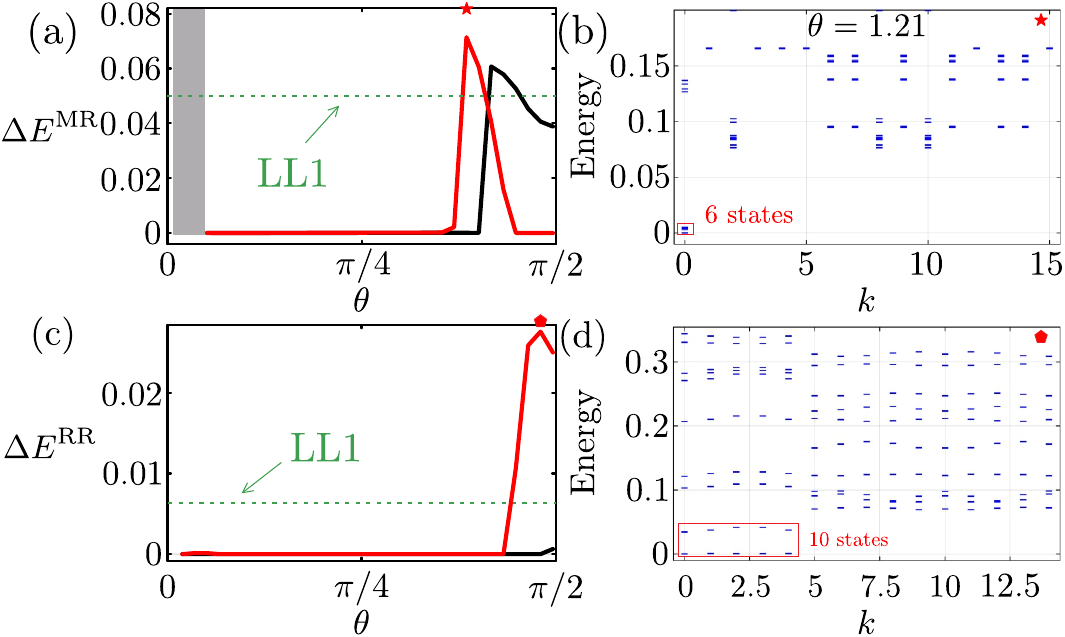}
\caption{Moore--Read and Read--Rezayi states from higher vortexable moir\'e bands. (a)~Many-body gap of the Moore--Read state, $\Delta E^{\mathrm{MR}}=E_7-E_6$, as a function of $\theta$ for an $N_s=16$ cluster. The red curve is obtained from two-band ED, whereas the black curve is obtained from single-band ED using only the first vortexable band. The green dashed line indicates the gap of the half-filled $n=1$ Landau level under the same screened Coulomb interaction with $d=0.27a$. Small-$\theta$ results show strong system-size dependence and are shown in gray. (b)~Many-body spectrum at $\theta=1.21$ and $\nu=3/2$ for an $N_s=16$ cluster using $d=0.27a$; the sixfold ground state manifold in the $k=0$ momentum sector is boxed. (c)~Many-body gap of the Read--Rezayi state,
$\Delta E^{\mathrm{RR}}=E_{11}-E_{10}$, as a function of $\theta$ for an $N_s=15$ cluster. The green dashed line indicates the gap of the three-fifths-filled first Landau level using the screened Coulomb interaction with a gate distance $d=0.69a$. (d)~Many-body spectrum at $\theta=1.51$ and $\nu=8/5$ for an $N_s=15$
cluster using $d=0.69a$; the tenfold ground state manifold in the momentum sectors $k=0,1,2,3,4$ is boxed. Panels (a) and (c) use a step size $\Delta\theta=0.05$.}
\label{fig:nonabelian}
\end{figure}

At filling $\nu=3/2$ for a system of $N_s=16$ unit cells, we find six nearly degenerate ground states in the $k=0$ momentum sector at $\theta=1.21$; see Fig.~\ref{fig:nonabelian}(b). To identify the nature of the ground state manifold, we compute its particle entanglement spectrum using a particle partition with $N_A=3$ and $N_B=21$ ($N_e=24$ and $N_s=16$). We find two gaps in the PES, above $560$ and $4912$ states; see Fig.~\ref{fig:pes_mr}(c). Note that the ground state at \(\theta=1.21\) can be adiabatically connected to a state obtained by completely filling band \(1\) and half filling band \(2\). If it is a Moore--Read (MR) state, its PES counting can be obtained using the ordinary Pauli principle for band $1$ and the generalized Pauli principle for band $2$, namely, no more than two electrons may occupy any four consecutive orbitals. For a particle cut with $N_A=3$, the expected total counting is obtained by summing over $N_A+1=4$ different ways of distributing the electrons between the two bands. The four ways are: (1) placing three electrons in band $1$, which gives $\binom{16}{3}=560$ states; (2) placing two electrons in band $1$ and one electron in band $2$, which gives
$\binom{16}{2}\times16=1920$ states; (3) placing one electron in band $1$ and two electrons in band $2$, which gives
$16\times\binom{16}{2}=1920$ states; and (4) placing three electrons in band $2$ subject to the generalized Pauli principle, which gives $512$ states. Together, these contributions sum to $4912$ states. The lowest $560$ states correspond to the possible configurations obtained by placing all three electrons in band $1$. The PES gap above $4912$ states supports the conclusion that the six ground states at $\theta=1.21$ are consistent with a Moore--Read state. We also studied an $N_s=12$ cluster at $\nu=3/2$ and observed a sixfold quasidegenerate ground state manifold, consistent with the expected momenta and ground state degeneracy of Moore--Read state from generalized Pauli principle~\cite{bernevig2012emergent}. The PES counting of this ground state manifold is also consistent with that of the Moore--Read state; see Fig.~\ref{fig:pes_mr}(a) and (b) in the Appendix. Moreover, within finite size resolution and under the same screened Coulomb interaction, the many-body gap of the Moore--Read state obtained from two-band ED at $\theta=1.21$ can be of the same order as that of the half-filled first Landau level, indicated by the green dashed line in Fig.~\ref{fig:nonabelian}(a).

For comparison, we also performed single-band ED projected onto the first vortexable band as a function of $\theta$, using the same gate distance $d\approx0.27a$. In the finite systems studied, the largest single-band gap occurs near $\theta\approx1.31$, whereas the two-band calculation gives its largest gap near $\theta\approx1.21$ (see Fig.~\ref{fig:nonabelian}(a)). Thus, within finite size resolution, including both bands moves the maximum of the Moore--Read gap toward smaller $\theta$. We further examine the stability of the Moore--Read state in the two-band calculation as a function of the gate distance $d$ and find that it remains the ground state for $0.1a\leq d\leq1.4a$, with the gap maximum occurring at $\theta\approx1.2\text{--}1.3$, depending on $d$; see Fig.~\ref{fig:mr_stability}(a). The many-body gap of the Moore--Read state increases with the gate distance because of the increased interaction range. The Moore--Read gap in the higher vortexable system begins to saturate near $d\sim0.69a$, whereas that of the half-filled first Landau level continues to increase over the range studied, up to $d\sim1.4a$; see Fig.~\ref{fig:mr_stability}(b).

At $\nu=1/2$, by contrast, we do not find a Moore--Read ground state anywhere in the range of $\theta$ studied. Instead, the ED spectrum shows low energy states with momenta and degeneracy that are consistent with composite Fermi liquid~\cite{dong2023composite}; see Fig.~\ref{fig:phasediagram}(b).

\subsection{\texorpdfstring{Read--Rezayi state at $\nu=8/5$}{Read--Rezayi state at nu=8/5}}
\label{sec:nu85}
We now turn to the Read--Rezayi state identified near $\theta=\pi/2$ in Sec.~\ref{sec:nu85_phases}. Within finite size resolution, the many-body gap of the Read--Rezayi state obtained from two-band ED can be of the same order as that of the three-fifths-filled first Landau level under the same screened Coulomb interaction; see Fig.~\ref{fig:ll1_nu35} for the many-body spectrum and PES of the ground states of the $n=1$ Landau level at $3/5$ filling. Relative to single-band ED, two-band calculations shift the optimal quantum geometry regime for stabilizing the Read--Rezayi state, as shown in Fig.~\ref{fig:nonabelian}(c). 

For both the Moore--Read and Read--Rezayi states, our simulations show that, within finite size resolution, two-band ED can stabilize non-Abelian phases with many-body gaps of the same order as those of the corresponding $n=1$ Landau level non-Abelian states under the same screened Coulomb interaction.
\section{Discussion}
\label{sec:discussion}
The findings of this paper, obtained for a homobilayer system under a moir\'e-periodic strain field, are expected to extend to other higher vortexable moir\'e systems, such as double twisted bilayer graphene and strained bilayer graphene, owing to the common origin of the flat bands across different moir\'e Hamiltonians~\cite{fujimoto2025higher,liu2025theory,wan2023nearly}. Even though our calculations are performed in the chiral limit, we expect these topological states to persist beyond the chiral limit, where these bands acquire small bandwidth, as they are protected by a finite many-body gap. Given recent advances in the precise control of twist angles in two-dimensional materials~\cite{tang2024chip} and in techniques for applying periodic strain fields~\cite{jiang2017visualizing,zhang2019magnetotransport,mao2020evidence,cho2021highly,zhang2024patternable,zhang2024enhancing,kim2023strain}, the experimental realization of such higher vortexable Hamiltonians is plausible in the near future. In two-dimensional van der Waals materials, interlayer tunneling is typically weak. Thus, higher vortexable systems realized by coupling two vortexable van der Waals systems are naturally expected to lie near the $\theta=0$ end of our phase diagram. Our results therefore suggest that these systems provide a promising platform for realizing integer and fractional exciton insulators at zero magnetic field. These phases can be experimentally detected through a quenched Hall resistance in counterflow measurements near $\nu=1$ and $\nu=1/3$~\cite{eisenstein2014exciton}.

\begin{acknowledgments}
This work was supported in part by Air Force Office of Scientific Research MURI FA9550-23-1-0334 and the Office of Naval Research MURI N00014-20-1-2479, and by the Gordon and Betty Moore Foundation Award N031710 (K.S.).
D.X. acknowledges support by the U.S. Department of Energy, Office of Basic Energy Sciences, under Contract No.~DE-SC0012509.
M.R. acknowledges the Brown Investigator Award, a program of the Brown Science Foundation, the University of Washington College of Arts and Sciences, and the Kenneth K. Young Memorial Professorship for support. T.C. acknowledges support by the U.S. Department of Energy, Office of Basic Energy Sciences, under Contract No.~DE-SC0025327. This research used resources of the National Energy Research Scientific Computing Center, a DOE Office of Science User Facility supported by the Office of Science of the U.S. Department of Energy under Contract No. DE-AC02-05CH11231 using NERSC awards BES-ERCAP0037104 and BES-ERCAP0037097 This work was facilitated through the use of advanced computational, storage, and networking infrastructure provided by the AI-core as well as the Hyak supercomputer system funded by the University of Washington Molecular Engineering Materials Center at the University of Washington (DMR-2308979). X.W. acknowledges support from the UW Thouless Fellowship in Physics. 
\end{acknowledgments}
 
\appendix
\section{Origin of flat band wave functions at critical values of moir\'e potential parameters}
\label{app:origin}
To understand the origin of exact flat bands in the moir\'e system described by the Hamiltonian in Eq.~\eqref{eq:D} of the main text, we start from the following Hamiltonian
\begin{equation}
    \mathcal{H}_{\text{v}}(\mathbf{r}) = \begin{pmatrix}
        0 & \mathcal{D}_{\text{v}}^\dagger(\mathbf{r})\\\mathcal{D}_{\text{v}}(\mathbf{r}) & 0
    \end{pmatrix},\, \mathcal{D}_{\text{v}}(\mathbf{r}) = -4\overline{\partial}^2+\tilde{A}(\mathbf{r}),
\end{equation}
where $z=x+iy$ is the complex coordinate, overline stands for complex conjugation, and $\tilde{A}(\mathbf{r}) = A_x(\mathbf{r})+iA_y(\mathbf{r})$. This model describes a two-dimensional material with a chiral (or sublattice) symmetric quadratic band touching, subject to a periodic moir\'e strain potential $\tilde{A}(\mathbf{r})$~\cite{wan2023topological}. If the strain field satisfies $\tilde{A}(\mathcal{C}_{6z} \mathbf{r}) = e^{-2\pi i/3}\tilde{A}(\mathbf{r})$ and $\tilde{A}(\mathcal{M}_x \mathbf{r}) = \overline{\tilde{A}(\mathbf{r})}$ (where $\mathcal{C}_{6z}$ and $\mathcal{M}_x$ are 6-fold rotation about out-of-plane axis $z$ and mirror reflection $x\rightarrow-x$, respectively), then the Hamiltonian has $p6mm$ symmetry; it satisfies $\mathcal{H}_{\text{v}}(\mathcal{C}_{6z}\mathbf{r}) = D(\mathcal{C}_{6z}) \mathcal{H}_{\text{v}}(\mathbf{r})D^\dagger(\mathcal{C}_{6z})$ with $D(\mathcal{C}_{6z}) = \text{diag}\{\omega,\omega^2\}$ and $\mathcal{H}_{\text{v}}(\mathcal{M}_{x}\mathbf{r}) = D(\mathcal{M}_x) \mathcal{H}_{\text{v}}(\mathbf{r})D^\dagger(\mathcal{M}_x)$ with $D(\mathcal{M}_x) = \sigma_x$. Furthermore, by construction, the Hamiltonian has chiral symmetry $\sigma_z \mathcal{H}_{\text{v}}(\mathbf{r})\sigma_z = -\mathcal{H}_{\text{v}}(\mathbf{r})$ and time reversal symmetry $\sigma_x \mathcal{H}_{\text{v}}^*(\mathbf{r})\sigma_x = \mathcal{H}_{\text{v}}(\mathbf{r})$. It was shown in~\cite{wan2023topological,eugenio2022twisted,sarkar2023symmetry} that this type of Hamiltonians hosts exact flat bands upon tuning control parameters that change $\tilde{A}(\mathbf{r})$. An exact flat band of $\mathcal{H}_{\text{v}}(\mathbf{r})$ at energy $E= 0$ with wave function $\Psi_\mathbf{k}(\mathbf{r})$ satisfies $\mathcal{H}_{\text{v}}(\mathbf{r})\Psi_\mathbf{k}(\mathbf{r}) = \mathbf{0}$ for all $\mathbf{k}$. The construction of such a $\Psi_\mathbf{k}(\mathbf{r})$ is as follows. Note that due to $\mathcal{C}_{6z}$ and chiral symmetry, the twofold degeneracy of the quadratic band crossing at $\Gamma$ point remains at $E= 0$ for any $\tilde{A}(\mathbf{r})$ that keeps $\mathcal{C}_{6z}$ symmetry. This means that there are always two sublattice polarized wave functions $\Psi_{\Gamma,1}(\mathbf{r}) = \{\psi_{\Gamma}(\mathbf{r}),\mathbf{0}\}$ and $\Psi_{\Gamma,2}(\mathbf{r}) = \{\mathbf{0},\psi_{\Gamma}^*(\mathbf{r})\}$ satisfying $\mathcal{H}_{\text{v}}(\mathbf{r})\Psi_{\Gamma,i}(\mathbf{r})=\mathbf{0}$, or equivalently $\mathcal{D}_{\text{v}}(\mathbf{r})\psi_{\Gamma}(\mathbf{r})=0$. If there are exact flat bands, the wave functions can be written as $\{\psi_\mathbf{k}(\mathbf{r}),\mathbf{0}\}$ and $\{\mathbf{0},\psi_{-\mathbf{k}}^*(\mathbf{r})\}$. Since the kinetic part of $\mathcal{D}(\mathbf{r})$ only contains antiholomorphic derivative, the trial wave function is naturally $\psi_{\mathbf{k}}(\mathbf{r}) = f_\mathbf{k}(z)\psi_{\Gamma}(\mathbf{r})$, where $f_\mathbf{k}(z)$ is a holomorphic function satisfying $\overline{\partial}f_\mathbf{k}(z) = 0$. The function $f_\mathbf{k}(z)$ needs to satisfy Bloch periodicity (translation by moir\'e lattice vector $\mathbf{a}$ gives phase shift $e^{i\mathbf{k}\cdot\mathbf{a}}$). However, from Liouville's theorem, such a holomorphic function must have poles, making $\psi_\mathbf{k}(\mathbf{r})$ divergent, unless $\psi_{\Gamma}(\mathbf{r})$ has a zero that cancels the pole. Conversely, if $\psi_{\Gamma}(\mathbf{r})$ has zero at $\mathbf{r}_0$, a Bloch periodic holomorphic function
\begin{equation}
\label{eq:fkz}
\begin{split}
    &f_\mathbf{k}(z;\mathbf{r}_0)  = e^{i (\mathbf{k}\cdot\mathbf{a}_1) z/a_1}\frac{\vartheta\left(\frac{z-z_0}{a_1}-\frac{k}{G_2},\tau\right)}{\vartheta\left(\frac{z-z_0}{a_1},\tau\right)}= e^{i\mathbf{k}\cdot\mathbf{r}} \tilde{f}_\mathbf{k}(\mathbf{r};\mathbf{r}_0),\\
    &\text{ where }\tilde{f}_\mathbf{k}(\mathbf{r};\mathbf{r}_0) = e^{-i(\mathbf{G}_2\cdot\mathbf{r})k/G_2}\frac{\vartheta\left(\frac{z-z_0}{a_1}-\frac{k}{G_2},\tau\right)}{\vartheta\left(\frac{z-z_0}{a_1},\tau\right)},
\end{split}
\end{equation}
with a pole at $\mathbf{r}_0$ can be constructed. Here $\vartheta (z,\tau) = -i\sum_{n=-\infty}^\infty(-1)^n e^{\pi i \tau(n+1/2)^2+\pi i(2n+1)z}$ is the Jacobi theta function of the first type~\cite{ledwith2020fractional}, $\mathbf{a}_i$ are lattice vectors, $\mathbf{G}_i$ are the corresponding reciprocal lattice vectors ($\mathbf{a}_i\cdot\mathbf{G}_j = 2\pi \delta_{ij}$), $a_i = (\mathbf{a}_i)_x+ i (\mathbf{a}_i)_y$, $G_i = (\mathbf{G}_i)_x+ i (\mathbf{G}_i)_y$, $z_0 = (\mathbf{r}_0)_x+ i (\mathbf{r}_0)_y$, $k = k_x + i k_y$, and $\tau = a_2/a_1$. Remarkably, at ``magic'' values of parameters of $\tilde{A}(\mathbf{r})$, the wave function $\psi_{\Gamma}(\mathbf{r})$ has a zero~\cite{wan2023topological}, which allows for such $f_\mathbf{k}(z;\mathbf{r}_0)$, and in turn gives rise to two exact flat bands.
Furthermore, the periodic part $\tilde{f}_\mathbf{k}(\mathbf{r};\mathbf{r}_0)$ is a holomorphic function of $k$: $\overline{\partial_{k}}\tilde{f}_\mathbf{k}(\mathbf{r};\mathbf{r}_0) = \frac{1}{2}[(\partial_{k_x}+i\partial_{k_y})]\tilde{f}_\mathbf{k}(\mathbf{r};\mathbf{r}_0) = 0$~\cite{kharchev2015theta,ledwith2020fractional}; this property along with the presence of the zero in the wave function can be used to prove that wave functions of this form carry Chern number $C = \pm 1$ (see~\cite{wang2021exact}). Moreover, since the periodic part $e^{-i\mathbf{k}\cdot\mathbf{r}}\psi_\mathbf{k}(\mathbf{r}) = \tilde{f}_\mathbf{k}(\mathbf{r};\mathbf{r}_0)\psi_\Gamma(\mathbf{r})$ is holomorphic in $k=k_x+ik_y$, this wave function satisfies ideal quantum geometry i.e., trace of quantum metric $g(\mathbf{k})$ equals the absolute value of the Berry curvature $F_{xy}(\mathbf{k})$ at all momenta $\mathbf{k}$ (see~\cite{ledwith2020fractional} for a proof). Lastly, the wave function $\psi_\mathbf{k}(\mathbf{r})$ can be written as
\begin{equation}
\begin{split}
    &\psi_\mathbf{k}(\mathbf{r}) = f_\mathbf{k}(z;\mathbf{r}_0)\psi_\Gamma(\mathbf{r}) = \psi_\mathbf{k}^\text{LLL}(\mathbf{r})h(\mathbf{r}),\\ 
    &\text{where }\psi_\mathbf{k}^\text{LLL}(\mathbf{r}) = e^{i (\mathbf{k}\cdot\mathbf{a}_1) z/a_1}\vartheta\left(\frac{z-z_0}{a_1}-\frac{k}{G_2},\tau\right) \\
    &\exp\left(- \frac{(\hat{G}_2\cdot(\mathbf{r}-\mathbf{r}_0))^2}{2\ell_B^2}\right)\text{ and }h(\mathbf{r})=\frac{\psi_\Gamma(\mathbf{r})}{\psi_\Gamma^\text{LLL}(\mathbf{r})},
\end{split}
\end{equation}
where $\psi_\mathbf{k}^\text{LLL}(\mathbf{r})$ is the $n=0$ Landau level wave function on a torus in the Landau gauge $\mathbf{A}(\mathbf{r}) = B_0(\hat{G}_2\cdot(\mathbf{r}-\mathbf{r}_0))(\hat{z}\times\hat{G}_2)$, where $\hat{G}_2 = \mathbf{G}_2/|\mathbf{G}_2|$ and magnetic flux per moir\'e unit cell is one flux quantum $B_0 \hat{z}\cdot(\mathbf{a}_1\times\mathbf{a}_2)=2\pi \hbar/e$ such that the magnetic length $\ell_B$ satisfies $2\pi \ell_B^2 = \hat{z}\cdot(\mathbf{a}_1\times\mathbf{a}_2)$. It can be verified using the definition of Jacobi theta function that $\psi_\mathbf{k}^\text{LLL}(\mathbf{r})$ satisfies magnetic Bloch-periodicity
\begin{equation}
\begin{split}
     \psi_\mathbf{k}^\text{LLL}(\mathbf{r}+\mathbf{a}_1) &= -e^{i\mathbf{k}\cdot\mathbf{a}_1}\psi_\mathbf{k}^\text{LLL}(\mathbf{r}),\\
     \,\psi_\mathbf{k}^\text{LLL}(\mathbf{r}+\mathbf{a}_2) &= -e^{i\mathbf{k}\cdot\mathbf{a}_2}e^{-2\pi i\frac{\mathbf{a}_1\cdot(\mathbf{r}-\mathbf{r}_0+\mathbf{a}_2/2)}{|\mathbf{a}_1|^2}}\psi_\mathbf{k}^\text{LLL}(\mathbf{r}).
\end{split}
\end{equation}
Due to this, along with the fact that $\psi_\Gamma(\mathbf{r}+\mathbf{a}_i) = \psi_\Gamma(\mathbf{r})$, we see that $\psi_\mathbf{k}(\mathbf{r}+\mathbf{a}_i) = e^{i\mathbf{k}\cdot\mathbf{a}_i}\psi_\mathbf{k}(\mathbf{r})$.

With the above knowledge, we next analyze the moir\'e Hamiltonian in the main text, 
\begin{equation}
\begin{split}
    \mathcal{H}(\mathbf{r}) &= \begin{pmatrix}0 & \mathcal{D}^\dagger(\mathbf{r})\\\mathcal{D}(\mathbf{r}) & 0\end{pmatrix},\,\\
    \mathcal{D}(\mathbf{r}) &= \begin{pmatrix}
        \mathcal{D}_{\text{v}}(\mathbf{r}) & 2 i \gamma\overline{\partial}\\0 & \mathcal{D}_{\text{v}}(\mathbf{r})
    \end{pmatrix}
\end{split}
\end{equation}
Clearly, if $\mathcal{D}_{\text{v}}(\mathbf{r})\psi_\mathbf{k}(\mathbf{r}) = 0$ for all $\mathbf{k}$, then $\mathcal{H}(\mathbf{r})$ must have an exact flat band at $E=0$ with wave function $\Psi_{\mathbf{k},1}(\mathbf{r})=\{\psi_\mathbf{k}(\mathbf{r}),0,0,0\}^T = \{\psi_\mathbf{k}^\text{LLL}(\mathbf{r}),0,0,0\}^Th(\mathbf{r})$. From our discussion on the properties of $\psi_\mathbf{k}(\mathbf{r})$, we know that this band must have Chern number $|C|=1$ and ideal quantum geometry. Next, we claim that $\mathcal{H}(\mathbf{r})$ has another flat band at $E=0$ with wave function 
\begin{equation}
\begin{split}
    &\Psi_{\mathbf{k},2}(\mathbf{r})= \{\ell_B\psi_\mathbf{k}^\text{LL1}(\mathbf{r})/\sqrt{8},\gamma^{-1}\psi_\mathbf{k}^\text{LLL}(\mathbf{r}),0,0\}^Th(\mathbf{r}) ,\\
    &\psi_\mathbf{k}^\text{LL1}(\mathbf{r}) = -i\sqrt{2}\ell_B e^{i (\mathbf{k}\cdot\mathbf{a}_1) z/a_1} e^{\left(- \frac{(\hat{G}_2\cdot(\mathbf{r}-\mathbf{r}_0))^2}{2\ell_B^2}\right)}\\
    &\left[\left(i \frac{\mathbf{k}\cdot\mathbf{a}_1}{a_1}-(\hat{G}_2\cdot(\mathbf{r}-\mathbf{r}_0))\frac{\overline{\hat{G}}_2}{\ell_B^2}\right)\vartheta\left(\frac{z-z_0}{a_1}-\frac{k}{G_2},\tau\right)\right.\\
    &\left.+\partial \vartheta\left(\frac{z-z_0}{a_1}-\frac{k}{G_2},\tau\right) \right],
\end{split}
\end{equation}
where $\overline{\hat{G}}_2 = ((\mathbf{G}_2)_x - i (\mathbf{G}_2)_y)/|\mathbf{G}_2|$, $\psi_\mathbf{k}^\text{LL1}(\mathbf{r})$ is $n=1$ Landau level wave function in the same Landau gauge mentioned earlier. Indeed $\psi_\mathbf{k}^\text{LL1}(\mathbf{r})$ written above is just $\psi_\mathbf{k}^\text{LL1}(\mathbf{r}) = a^\dagger \psi_\mathbf{k}^\text{LLL}(\mathbf{r})$, where $a^\dagger = \frac{\ell_B}{\sqrt{2}\hbar}(\Pi_x -i\Pi_y)$ is the Landau level ladder operator with $\Pi_\alpha = \hbar(-i\partial_\alpha+eA_\alpha/\hbar)$. Since $a^\dagger$ commutes with magnetic translation operation, $\psi_\mathbf{k}^\text{LL1}(\mathbf{r})$ satisfies the same magnetic Bloch periodicity as $\psi_\mathbf{k}^\text{LLL}(\mathbf{r})$; hence $\Psi_{\mathbf{k},2}(\mathbf{r})$ satisfies Bloch periodicity: $\Psi_{\mathbf{k},2}(\mathbf{r}+\mathbf{a}_i) = e^{i\mathbf{k}\cdot\mathbf{a}_i}\Psi_{\mathbf{k},2}(\mathbf{r})$. Next, to prove $\mathcal{H}(\mathbf{r})\Psi_{\mathbf{k},2}(\mathbf{r}) = \mathbf{0}$, we must show $\mathcal{D}(\mathbf{r})\{\ell_B\psi_\mathbf{k}^\text{LL1}(\mathbf{r})/\sqrt{8},\gamma^{-1}\psi_\mathbf{k}^\text{LLL}(\mathbf{r})\}^Th(\mathbf{r}) = \mathbf{0}$. Using 
$\mathcal{D}_{\mathrm{v}}(\mathbf{r})
\big[\psi_{\mathbf{k}}^{\mathrm{LL1}}(\mathbf{r})h(\mathbf{r})\big]
=
-\frac{4\sqrt{2}i}{\ell_B}\,
\overline{\partial}\,
\big[\psi_{\mathbf{k}}^{\mathrm{LLL}}(\mathbf{r})h(\mathbf{r})\big]
$ together with $\mathcal{D}_{\mathrm{v}}(\mathbf{r})
\big[\psi_{\mathbf{k}}^{\mathrm{LLL}}(\mathbf{r})h(\mathbf{r})\big]
=0$,
one directly obtains
\(\mathcal{H}(\mathbf{r})\Psi_{\mathbf{k},2}(\mathbf{r})=\mathbf{0}\).
Note that $\Psi_{\mathbf{k},1}(\mathbf{r})$ and $\Psi_{\mathbf{k},2}(\mathbf{r})$ are independent but not orthogonal to each other (they would be orthogonal if $h(\mathbf{r})=1$). We can construct the orthonormalized wave functions $\tilde{\Psi}_{\mathbf{k},1}(\mathbf{r})=\mathcal{N}_{\mathbf{k},1}\Psi_{\mathbf{k},1}(\mathbf{r})$ and $\tilde{\Psi}_{\mathbf{k},2}(\mathbf{r})=\mathcal{N}_{\mathbf{k},2}(\Psi_{\mathbf{k},2}(\mathbf{r}) - \langle\tilde{\Psi}_{\mathbf{k},1}|\Psi_{\mathbf{k},2} \rangle \tilde{\Psi}_{\mathbf{k},1}(\mathbf{r}))$, where $\mathcal{N}_{\mathbf{k},i}$ are normalization factors. 
\section{Cluster geometries used in exact diagonalization}
\label{app:clusters}
\begin{figure}[h]
\centering
\includegraphics[width=\columnwidth]{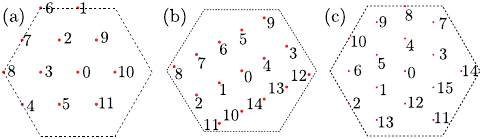}
\caption{The $12$-unit-cell cluster, $15$-unit-cell cluster, and $16$-unit-cell cluster used in exact diagonalization are shown in (a)--(c), respectively.}
\label{fig:clusters}
\end{figure}

Figure~\ref{fig:clusters} shows the $12$-, $15$-, and $16$-unit-cell clusters used in the exact diagonalization calculations throughout this work.
 
\section{Adiabatic continuity of the \texorpdfstring{$\theta=\pi/2$}{theta=pi/2} ground states}
\label{app:adiabatic}
\subsection{$\nu=1$}
\begin{figure}[h]
\centering
\includegraphics[width=\columnwidth]{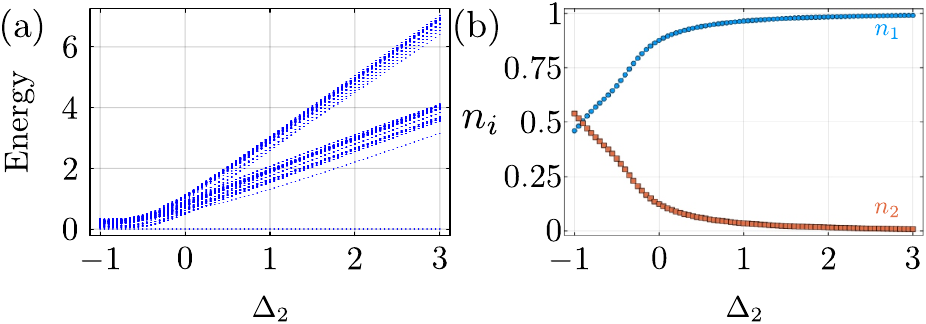}
\caption{
Energy spectra and occupation numbers at \(\nu=1\) and
\(\theta=\pi/2\) for an \(N_s=12\) cluster as a function of the
single particle energy \(\Delta_2\) of the first vortexable band, obtained from
multiband ED with
\(H=H_{\mathrm{int}}+\Delta_2\sum_{\mathbf{k}}
c^{\dagger}_{\mathbf{k},2}c_{\mathbf{k},2}\). (a)~Many-body energies versus $\Delta_{2}$. (b)~Ground state occupation numbers $n_{i}=\tfrac{1}{N_{s}}\sum_{\mathbf{k}}\langle c^{\dagger}_{\mathbf{k},i}c_{\mathbf{k},i}\rangle$ versus $\Delta_{2}$.}
\label{fig:adiabatic}
\end{figure}

The energy spectrum at $\nu=1$ and $\theta=\pi/2$ shows one gapped ground state at $k=0$ (Fig.~\ref{fig:exciton}(b)) with many-body Chern number $C_{\mathrm{mb}}=1$. The occupation numbers show that the electrons are not fully polarized in either band; moreover, single-band ED that completely fills band $1$ yields a ground state energy of $1.28 U$, different from the value $1.12 U$ obtained from multiband ED under the same screened Coulomb interaction. We nonetheless show that this state, although not fully polarized and not a single Slater determinant state, is adiabatically connected to a Chern insulator obtained by fully filling band $1$ (see Fig.~\ref{fig:adiabatic}). The adiabatic path is realized by adding a single particle energy $\Delta_{2}$ to band $2$, i.e., $H=H_{\mathrm{int}}+\Delta_{2}\sum_{\mathbf{k}}c^{\dagger}_{\mathbf{k},2}c_{\mathbf{k},2}$. As $\Delta_{2}$ increases, the electrons become fully polarized into band $1$ while the many-body gap increases monotonically.
\subsection{$\nu=2/5$}
\begin{figure}[h]
\centering
\includegraphics[width=\columnwidth]{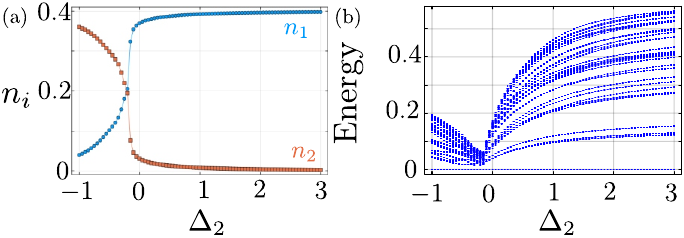}
\caption{Identification of the ground state at $\theta=\pi/2$ and $\nu=2/5$ for an \(N_s=15\) cluster. (a)~Band occupation numbers averaged over the five ground states versus
\(\Delta_2\). (b)~Many-body energy spectrum versus \(\Delta_2\) for
\(H=H_{\mathrm{int}}+\Delta_2\sum_{\mathbf{k}}
c^{\dagger}_{\mathbf{k},2}c_{\mathbf{k},2}\).}
\label{fig:pol_nu25}
\end{figure}

The energy spectrum at \(\nu=2/5\) and \(\theta=\pi/2\) shows one gapped ground state in each of the five center of mass momentum sectors \(k=0,1,2,3,4\), forming a fivefold quasidegenerate ground state manifold with average many-body Chern number \(C_{\mathrm{mb}}=2/5\). The occupation numbers show that the electrons are not fully polarized in either band (see Fig.~\ref{fig:pol_nu25}(a) at $\Delta_2=0$).
As \(\Delta_2\) increases, the electrons become fully polarized into band \(1\) near \(\Delta_2=3\), while the many-body gap separating the fivefold ground states from the excited states remains open; see Fig.~\ref{fig:pol_nu25}(b). The \(\theta=\pi/2\) state is therefore adiabatically connected to the \(2/5\) Jain state, which corresponds to electrons polarized into band $1$ around \(\Delta_2=3\).

\section{Entanglement spectra}
\label{app:pes}
This appendix collects the detailed particle/hole entanglement spectra and related PES counting used to identify the ground states throughout the main text.

\subsection{Large $\theta$ at $\nu=1/3$ }
\begin{figure}[h]
\centering
\includegraphics[width=0.8\columnwidth]{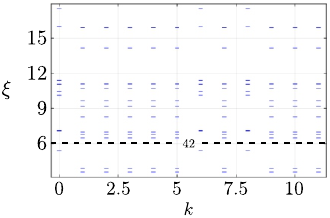}
\caption{PES of the \(\nu=1/3\) ground state manifold at
\(\theta=\pi/2\) for \(N_s=12\), \(N_A=2\), and \(d=0.27a\); see
Fig.~\ref{fig:exciton}(d) for the many-body spectrum. The PES gap above
\(42\) states agrees with the \(1/3\)-Laughlin generalized Pauli principle---no more than one electron may occupy any three consecutive
orbitals.}
\label{fig:pes_nu13}
\end{figure}
For \(N_A=2\) and \(N_s=12\), the PES of the \(\nu=1/3\) ground state manifold at \(\theta=\pi/2\) exhibits a gap above \(42\) states, consistent with the generalized Pauli principle that no more than one electron may occupy any three consecutive orbitals; see Fig.~\ref{fig:pes_nu13}.
 
\subsection{Small $\theta$ at $\nu=2/5$}
\begin{figure}[h]
\centering
\includegraphics[width=\columnwidth]{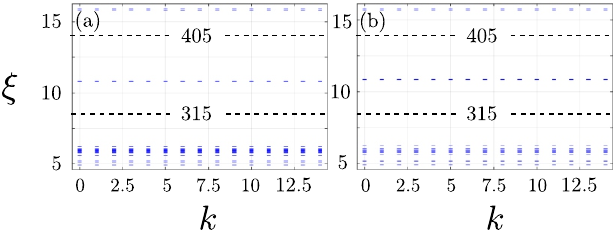}
\caption{Identification of the ground state at $\theta=0.05$ and $\nu=2/5$. (a)~PES of the five ground states at $\theta=0.05$ for the $15$-unit-cell cluster with $N_{A}=2$. (b)~PES of the Halperin-$332$ state from a quantum Hall bilayer at $\nu=2/5$.}
\label{fig:pes_nu25}
\end{figure}
We identify the \(\theta=0.05\) ground state manifold as a Halperin-\(332\) state by comparing its PES, which exhibits two gaps above \(315\) and \(405\) states, with that of a Halperin-\(332\) state in a quantum Hall bilayer at \(\nu=2/5\); see Fig.~\ref{fig:pes_nu25}.
\subsection{Small $\theta$ at $\nu=8/5$}
\begin{figure}[h]
\centering
\includegraphics[width=\columnwidth]{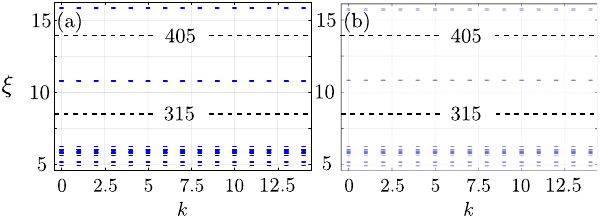}
\caption{Identification at $\theta=0.05$ and $\nu=8/5$. (a)~Hole PES of the five ground states for the $15$-unit-cell cluster with $N_{A}=2$. (b)~PES of the Halperin-$332$ state from a quantum Hall bilayer at $\nu=2/5$. Both show gaps above $315$ and $405$ states.}
\label{fig:pes_nu85_small}
\end{figure}

The hole-PES gaps of the fivefold ground state manifold agree with the PES gaps of a Halperin-\(332\) state in a quantum Hall bilayer at \(2/5\) filling; see Fig.~\ref{fig:pes_nu85_small}. This identifies the small-\(\theta\) ground state manifold as a Halperin-\(332\) state of holes.
\subsection{Intermediate $\theta$ at $\nu=8/5$}
\begin{figure}[h]
\centering
\includegraphics[width=\columnwidth]{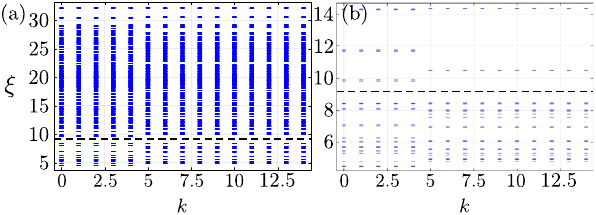}
\caption{(a)~Hole PES of the five ground states at \(\theta=0.25\) and \(\nu=8/5\) for the $15$-unit-cell cluster with $N_{A}=3$. (b)~PES of the $2/5$ Jain state from the LLL at $\nu=2/5$ with $N_{A}=3$. Both dashed lines correspond to a gap above $350$ states.}
\label{fig:pes_nu85_large}
\end{figure}

The hole PES of the ground state manifold at \(\theta=0.25\) exhibits a gap above \(350\) states, which is also present in the PES of a \(2/5\) Jain state in the LLL; see Fig.~\ref{fig:pes_nu85_large}.
\subsection{$\nu=3/2$ Moore--Read state at $\theta=1.21$}
The multiband Moore--Read state at \(\nu=3/2\) can be adiabatically
connected to a state obtained by completely filling band \(1\) and
half-filling band \(2\). For a particle cut containing \(N_A\) particles, the particles can be
distributed between the two bands according to
\(N_A=N_A^{(1)}+N_A^{(2)}\). Band \(1\) contributes the ordinary fermionic
counting \(\binom{N_s}{N_A^{(1)}}\), while band \(2\) contributes the
Moore--Read counting
\(\mathcal{N}_{\mathrm{MR}}(N_s,N_A^{(2)})\), obtained from the generalized
Pauli principle that no more than two particles may occupy any four
consecutive orbitals. Summing over all ways of distributing the particles
between the two bands gives $\mathcal{N}^\text{MR}_{\mathrm{PES}}
=
\sum_{N_A^{(1)}+N_A^{(2)}=N_A}
\binom{N_s}{N_A^{(1)}}
\mathcal{N}_{\text{MR}}\!\left(N_s, N_A^{(2)}\right)$.
\begin{figure}[t]
\centering
\includegraphics[width=\columnwidth]{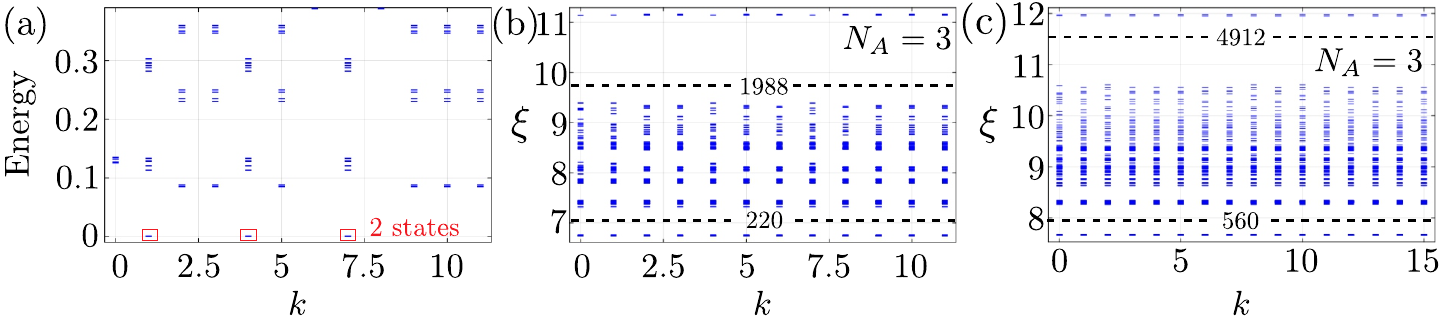}
\caption{(a)~Many-body spectrum at $\nu=3/2$ and $\theta=1.21$ with $d=0.27a$ on the $12$-unit-cell cluster, with two nearly degenerate ground states at each of $k=1,4,7$. (b)~PES of the six ground states in (a) with $N_{A}=3$, showing gaps above $220$ and $1988$ states. (c)~PES of the sixfold ground state manifold at \(\theta=1.21\) and \(d=0.27a\) for the \(N_s=16\) cluster with \(N_A=3\), showing gaps above \(560\) and \(4912\) states.}
\label{fig:pes_mr}
\end{figure}
For the \(N_s=12\) cluster, we find two quasidegenerate ground states in each of the momentum sectors \(k=1,4,7\), giving a total of six states; see Fig.~\ref{fig:pes_mr}(a). Their \(N_A=3\) PES exhibits gaps above \(220\) and \(1988\) states; see Fig.~\ref{fig:pes_mr}(b). The number $220$ is consistent with the number of possible ways of putting three electrons in band $1$. The number \(1988\) is consistent with the sum over the following four ways of distributing the three electrons between the two bands: (1) putting three electrons in band $1$, which gives $\binom{12}{3}=220$ states; (2-3) putting one electron in band $1$ and two electrons in band $2$, and vice versa, which gives $2\times12\times 66=1584$; (4) putting all three electrons in band $2$ following generalized Pauli principle (no more than two electrons within four consecutive orbitals), which gives $184$. Together they sum to 1988, the number of states below the second gap in Fig.~\ref{fig:pes_mr}(b). In both finite size systems we studied, the ground state remains consistent with Moore--Read state. 

\subsection{$\nu=8/5$ Read--Rezayi state at large $\theta$}
\begin{figure}[h]
\centering
\includegraphics[width=0.8\columnwidth]{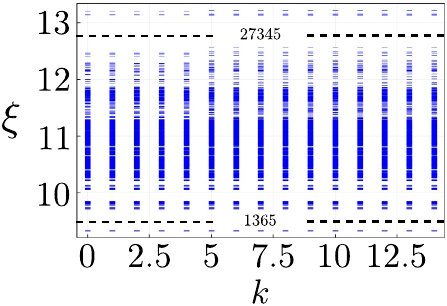}
\caption{PES of the tenfold Read--Rezayi ground state manifold at
\(\nu=8/5\) and \(\theta=1.51\) for an \(N_s=15\) cluster with \(N_A=4\) and \(d=0.69a\), showing gaps above \(1365\) and \(27345\) states; see Fig.~\ref{fig:nonabelian}(d) for the many-body spectrum.}
\label{fig:pes_rr}
\end{figure}

\begin{figure}[h]
\centering
\includegraphics[width=\columnwidth]{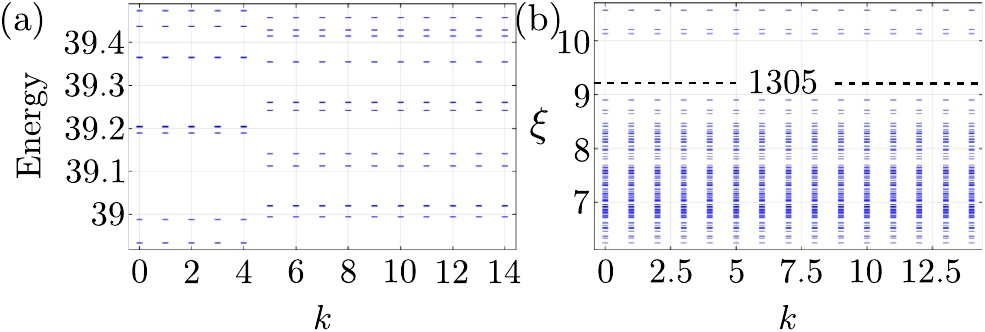}
\caption{(a)~Many-body spectrum of the first Landau level at $\nu=3/5$ for an $N_s=15$ cluster with screened Coulomb interaction and $d=0.69a$. (b)~PES of the ten ground states, with two states in each momentum sector
\(k=0,1,2,3,4\), using \(N_A=4\). The PES exhibits a gap above \(1305\)
states, consistent with the single-band Read--Rezayi PES counting.}
\label{fig:ll1_nu35}
\end{figure}

Similar to the multiband Moore--Read state at \(\nu=3/2\), the expected
total PES counting of the multiband Read--Rezayi state is obtained by
summing over all ways of distributing the \(N_A\) particles between the
two bands:
$
\mathcal{N}^{\mathrm{RR}}_{\mathrm{PES}}
=
\sum_{N_A^{(1)}+N_A^{(2)}=N_A}
\binom{N_s}{N_A^{(1)}}
\mathcal{N}_{\mathrm{RR}}\!\left(N_s,N_A^{(2)}\right)
$, where $\mathcal{N}_{\mathrm{RR}}\!\left(N_s, N_A^{(2)}\right)$ is the number of ways of putting $N_A^{(2)}$ particles following the generalized Pauli principle---no more than three electrons per five consecutive orbitals. For \(N_A=4\), the PES exhibits gaps above \(1365\) and \(27345\) states, whose counting is detailed in the main text; see Fig.~\ref{fig:pes_rr}. For comparison, we show the many-body spectrum of RR state from first Landau level under the same screened Coulomb interaction at $\nu=3/5$ in Fig.~\ref{fig:ll1_nu35}(a) with its PES shown in Fig.~\ref{fig:ll1_nu35}(b). There is a gap above $1305$ states in the PES which matches the PES counting of single band Read--Rezayi state of $N_A=4$ particles in $N_s=15$ orbitals on the torus.

\subsection{$\nu=2/3$ at both $\theta$ limits}
The many-body spectra at \(\nu=2/3\) for \(\theta=0.05\) and
\(\theta=\pi/2\) exhibit the same ground state degeneracy and momentum
sectors; see Fig.~\ref{fig:spectra_2325}. Nevertheless, their entanglement
spectra reveal different underlying phases.

\begin{figure}[h]
\centering
\includegraphics[width=\columnwidth]{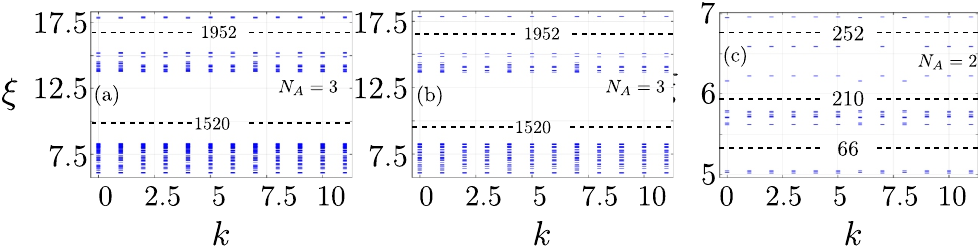}
\caption{Entanglement spectra at \(\nu=2/3\) for an \(N_s=12\) cluster. (a)~PES of the threefold ground state manifold at \(\theta=0.05\) with \(N_A=3\), showing gaps above \(1520\) and \(1952\) states. (b)~PES of the Halperin-\(112\) state obtained from a quantum Hall bilayer at \(\nu=2/3\) with \(N_A=3\), showing the same gaps as in (a). (c)~Hole PES of the threefold ground state manifold at \(\theta=\pi/2\) with \(N_A=2\), showing gaps above \(66\), \(210\), and \(252\) states.}
\label{fig:pes_nu23}
\end{figure}

At \(\theta=0.05\), the PES with \(N_A=3\) exhibits two gaps, above \(1520\)
and \(1952\) states; see Fig.~\ref{fig:pes_nu23}(a). These gaps agree
with those of the Halperin-\(112\) state in a quantum Hall bilayer at
\(\nu=2/3\), shown in Fig.~\ref{fig:pes_nu23}(b). Therefore, at \(\theta=0.05\), we identify the ground state manifold as a Halperin-\(112\) state.

At \(\theta=\pi/2\), the hole
PES with \(N_A=2\) exhibits gaps above \(66\), \(210\), and \(252\)
states, consistent with the multiband \(2/3\)-FCI counting detailed in
the main text; see Fig.~\ref{fig:pes_nu23}.
\section{Gap closing at $\nu=8/5$}
\label{app:gapclose_nu85}
\begin{figure}[h]
\centering
\includegraphics[width=\columnwidth]{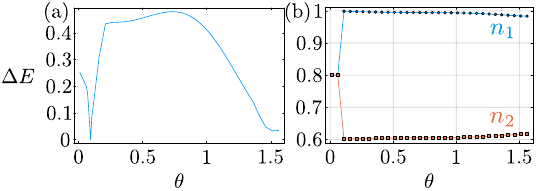}
\caption{Gap closing at \(\nu=8/5\) for an \(N_s=15\) cluster with
\(d=0.69a\). (a)~Many-body gap
\(\Delta E=E_6-E_5\) versus \(\theta\).
(b)~Band occupation numbers averaged over the lowest five states versus
\(\theta\).}
\label{fig:gapclose_nu85}
\end{figure}

For small $\theta$, the ground state at $\nu=8/5$ is the Halperin-$332$ state of holes identified in Sec.~\ref{sec:nu85_phases}; near $\theta\approx0.1$ the many-body gap closes and the band occupation numbers exhibit a jump (see Fig.~\ref{fig:gapclose_nu85}).
\section{Stability of the Moore--Read state}
\label{app:moore_read_stability}
\begin{figure}[h]
\centering
\includegraphics[width=\columnwidth]{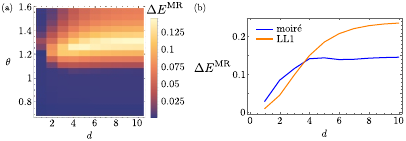}
\caption{(a)~Many-body gap of the Moore--Read state at $\nu=3/2$ as a function of $d$ and $\theta$. (b)~Blue: maximum many-body gap of the Moore--Read state at $\nu=3/2$ over $\theta$ for different $d$. Orange: many-body gap of the half-filled first Landau level Moore--Read state at $\nu=1/2$ for different $d$. Both panels use a $12$-unit-cell system with moir\'e lattice constant $a=4\pi/\sqrt{3}$. This is equivalent to setting $G=1$.}
\label{fig:mr_stability}
\end{figure}
The $(\theta,d)$ dependence of the Moore--Read gap and its comparison with the half-filled first Landau level gap are shown in Fig.~\ref{fig:mr_stability}. The range of $\theta$ over which the Moore--Read state is stabilized grows as $d$ increases from $d=1\approx0.14a$ to $d=5\approx0.69a$ and then saturates. At the gate distance $d\approx0.27a$ used in Sec.~\ref{sec:nu32}, the maximum two-band gap occurs near $\theta\approx1.2$. Across all values of $d$ studied here, the maximizing angle remains within the range $\theta\approx1.2\text{--}1.3$.
\bibliographystyle{apsrev4-2}
\bibliography{ref}

\end{document}